\documentclass[a4paper,11pt]{article}

\usepackage{xparse}
\NewDocumentCommand \deq { o m }{
\begin{equation}
#2
\IfNoValueF{#1}{\label{#1}}
\end{equation}
}

\usepackage{tikz,tikz-cd}
\usetikzlibrary{arrows}

\usepackage{amsmath,amsfonts,amssymb,amsthm,mathrsfs}
\usepackage[vcentermath]{youngtab}
\usepackage{graphicx,graphics}
\graphicspath{{image/}}
\usepackage{enumitem}
\usepackage{cases}
\usepackage{epsfig}
\usepackage{latexsym} 
\usepackage{hyperref}
\usepackage{xcolor}
\hypersetup{
    colorlinks,
    linkcolor={red!30!black},
    citecolor={blue!40!black},
    urlcolor={blue!80!black}
}
\usepackage{amsmath,amssymb,mathtools}
\usepackage{pdfpages} 
\usepackage{tikz}
\usetikzlibrary{trees,arrows}
\tikzset{level 1/.style={level distance=1.5cm, sibling distance=3.5cm}}
\tikzset{level 2/.style={level distance=1.5cm, sibling distance=2cm}}
 
\usepackage{pdfpages}
\usepackage[left=2.5cm,top=2.3cm,right=2.5cm]{geometry}

\usepackage{physics}

\DeclareMathOperator{\Spin}{Spin}
\DeclareMathOperator{\SO}{SO}
\DeclareMathOperator{\SU}{SU}

\DeclareMathOperator{\graph}{graph}

\newsavebox{\uuunit}
\sbox{\uuunit}
    {\setlength{\unitlength}{0.825em}
     \begin{picture}(0.6,0.7)
        \thinlines
        \put(0,0){\line(1,0){0.5}}
        \put(0.15,0){\line(0,1){0.7}}
        \put(0.35,0){\line(0,1){0.8}}
       \multiput(0.3,0.8)(-0.04,-0.02){10}{\rule{0.5pt}{0.5pt}}
     \end {picture}}

\newtheorem{theorem}{Theorem}
\newtheorem{fact}[theorem]{Fact}
\theoremstyle{definition}
\newtheorem{dfn}[theorem]{Definition}

\DeclareMathOperator{\stab}{stab}
\def\lie#1{\ensuremath{\mathfrak{#1}}}
\def\O{\ensuremath{\mathscr{O}}}
\def\K{\ensuremath{\mathscr{K}}}
\def\bu{\ensuremath{\bullet}}
\def\L{\ensuremath{\mathscr{L}}}
\def\E{\ensuremath{\mathscr{E}}}
\def\lbr{\left[\!\left[}
\def\rbr{\right]\!\right]}
\def\PV{\ensuremath{\text{PV}}}
\def\eps{\ensuremath{\epsilon}}
\def\bdy{\ensuremath{\partial}}
\def\hol{\ensuremath{\text{hol}}}

\newcommand{\ba}{\begin{eqnarray*}}
\newcommand{\ea}{\end{eqnarray*}}
\newcommand{\ban}{\begin{eqnarray}}
\newcommand{\ean}{\end{eqnarray}}

\newcommand{\zbar}{\overline{z}}

\newcommand{\del}{\partial}

\def\N{\ensuremath{\mathscr{N}}}
\def\C{\ensuremath{\mathbb{C}}}
\def\Z{\ensuremath{\mathbb{Z}}}
\def\R{\ensuremath{\mathbb{R}}}

\def\H{\ensuremath{\mathbb{H}}}

\DeclareMathOperator{\Ext}{Ext}

\DeclareMathOperator{\Sym}{Sym}
\DeclareMathOperator{\ad}{ad}

\def\beq{\begin{equation}}
\def\bee{\begin{equation}}
\def\eeq{\end{equation}}
\def\bea{\begin{eqnarray}}
\def\eea{\end{eqnarray}}
\def\bd{\begin{displaymath}}
\def\ed{\end{displaymath}}

\title{\textbf{Holomorphic boundary conditions for topological field theories via branes in twisted supergravity}}
\author{Ilka Brunner\footnote{ilka.brunner@physik.uni-muenchen.de} \and Ioannis Lavdas\footnote{ioannis.lavdas@physik.uni-muenchen.de}\and Ingmar Saberi\footnote{i.saberi@physik.uni-muenchen.de}}
\date{%
\center{\textit{\small 
Arnold Sommerfeld Center for Theoretical Physics \\ Ludwig-Maximilians-Universit{\"a}t M\"unchen \\ Theresienstra\ss{}e 37 \\ 80333 M{\"u}nchen \\ Germany}}
}

\begin{document}

\maketitle

\begin{center}
\textbf{Abstract}
\end{center}
Three-dimensional $\N=4$ supersymmetric field theories admit a natural class of chiral half-BPS boundary conditions that preserve $\N=(0,4)$ supersymmetry.
 While such boundary conditions are not compatible with topological twists, 
deformations that define boundary conditions for the topological theories 
were recently introduced by Costello and Gaiotto.
Not all $\N=(0,4)$ boundary conditions admit such deformations. 
We revisit this construction, working directly in the setting of the holomorphically twisted theory 
and viewing the topological twists as further deformations. 
Properties of the construction are explained both purely in the context of 
holomorphic field theory and also by engineering the holomorphic
theory on the worldvolume of a D-brane. Our brane engineering approach combines the
intersecting brane configurations of Hanany--Witten with recent work of Costello
and Li on twisted supergravity. The latter approach allows to realize
holomorphically and topologically twisted field theories directly as worldvolume theories in deformed
supergravity backgrounds, and we make extensive use of this.

\vfill
\leftline{Preprint LMU-ASC 41/21}

\newpage 
\tableofcontents
\newpage

\section{Introduction}
\label{Intro}

Topological twists of supersymmetric field theories have been studied extensively over the past decades. One reason for this is that  twists  single out simple, often exactly solvable, subsectors of the initial physical theory. Performing a topological twist implies at least two different operations. First, the notion of Lorentz symmetry is modified by choosing a homomorphism to the global $R$-symmetry group. As a result,  one of the supercharges becomes a scalar, and hence squares to zero. In addition, one then takes the cohomology of this scalar supercharge. If the stress tensor becomes trivial in cohomology, the twist leads to a topological theory, but for many purposes it is interesting to also consider more general versions of twists. Such generalized twists have received much attention lately. For one thing, many more theories admit partial twists than can be topologically twisted. In addition, more of the structure of the initial theory survives, so that partial twists are ``less forgetful'' and provide more detailed information about the original theory. In particular, holomorphic twists play an important role, as holomorphicity still leads to strong constraints and solvable subsectors. Furthermore, even when topological twists exist, it is often quite useful to regard them as deformations of a holomorphic twist. 

In this paper, we study theories with boundaries, starting at level of the holomorphically twisted theory. In this setting, all half-BPS boundary conditions are compatible with the holomorphic supercharge, and give rise to holomorphic boundary conditions that can be described explicitly. We subsequently consider deformations of the holomorphic supercharge that lead to a topological twist of the bulk theory. Not all holomorphic boundary conditions will be compatible with such a deformation. Moreover, it  is not immediate that the further twist will  lead to a topological theory on the boundary condition. Indeed, in many examples, natural holomorphic boundary conditions for topological field theories can be constructed in this manner.

Precisely such holomorphic boundary conditions were considered together with a topological bulk in \cite{CostelloGaiotto,CostelloGaiottoCreutzig,CostelloDimofteGaiotto}, in the setting of three-dimensional field theories with $\N=4$ supersymmetry. Starting from there, an intricate structure was unraveled, where vertex operator algebras appear at special, holomorphic boundaries. This lead the authors of \cite{CostelloGaiotto,CostelloGaiottoCreutzig,CostelloDimofteGaiotto} to propose  a derived version of the relation between 3d Chern--Simons theory and rational conformal field theories. A number of conjectures arising from this basic insight were proposed and checked in detail. (A framework using equivariant factorization algebras that leads to rigorous proofs of some of the conjectures of~\cite{CostelloGaiotto} is presented in~\cite{Butson1,Butson2}.)

To summarize a few main points of this work: In the new, derived version of the correspondence, holomorphic vertex operator algebras arise as the cohomology of supercharges and are generically non-rational.
Modules of the VOA find  a natural home in this correspondence, since one expects that line operators ending at the boundary come equipped with  an action of the boundary operators. More precisely, one expects  a functor from the category of modules of VOAs (associated to certain boundary conditions) to the category of bulk line operators. This functor relates the identity line with the vacuum module of the VOA, a relation that is explored in several examples in~\cite{CostelloGaiottoCreutzig}. The simplest ones  are the free hypermultiplet, where  $\N=(0,4)$ boundary conditions are provided by setting one half of the fermions to zero at the boundary. By supersymmetry, this corresponds to all Dirichlet or all Neumann boundary conditions on the bosons. The corresponding VOAs are those of the symplectic fermion and boson, respectively. More interesting examples as well as general arguments lead to an interplay between properties of classes of VOAs as well as their representations and bulk topological field theories. This vast range of applications and consequences makes it clear that it is of interest to understand precisely how holomorphic boundaries for topological bulk theories arise, and how we can construct them.

As pointed out above, one possibility is to begin by studying boundary conditions in the holomorphic theory, and then to obtain the topological theory as a deformation. One then asks that the boundary conditions are compatible with this deformation. In this paper, we pursue this approach from various points of view. This includes a pure field theory analysis, as well as a realization of the theories as world volume theories on D-branes. 
While our methods are general, in this paper we mainly focus on the case of $\N=4$ supersymmetric field theories in three dimensions. 
The ingredients of the field-theoretic approach involve holomorphic twists of the vector multiplet and hypermultiplet, which were known  in  previous literature; we review them concretely. 
For the latter perspective, we embed three-dimensional $\N=4$ theories into type IIB string theory using the construction in \cite{HananyWitten}. There, a three-dimensional supersymmetric theory is obtained by suspending D3 branes between NS5 or D5 branes. This yields theories of free vector or free hypermultiplets, respectively. Further branes can be introduced to realize additional matter multiplets and to engineer more general gauge theories.

Different approaches make different features of the topological twist more transparent. In the context of geometric engineering, many field-theoretic constructions gain a pleasing interpretation in terms of the geometry of the corresponding target space. For example, purely geometric constructions allow one to twist the action of Lorentz symmetry, by embedding a brane worldvolume theory along a calibrated cycle in a nontrivial supergravity background~\cite{BershadskySadovVafa}.
It is natural to ask whether or not the other part of the twisting procedure can be made manifest in this setting by understanding it in terms of a different choice of supergravity background.

It turns out  that there is a  natural interpretation of twisting a brane worldvolume theory in terms of a coupling to a particular supergravity background. This interpretation was developed in the work of Costello and Li~\cite{CostelloLi}. They point out that the supergravity theory contains bosonic ghost fields, corresponding to gauging  local supersymmetry. If the metric background admits Killing spinors, these ghosts may acquire vacuum expectation values, which  have the effect of adding the corresponding preserved supercharge to the BRST differential.  A brane worldvolume theory in such  a background is thus automatically ``twisted'' in the second  sense discussed above.
One can therefore geometrically engineer holomorphically  twisted theories directly by considering \emph{twisted supergravity}. Our second approach will be to construct some examples of this holomorphic version of the Hanany--Witten construction.

In the field theory part, we give a formulation of the procedure of~\cite{CostelloGaiotto} at the level of the holomorphic twist. Instead of constructing boundaries and their deformations on the level of the initial, physical theory and deforming them at the level of the action, we start with a holomorphically twisted theory and employ the BV formalism.  
We then consider topological twists as deformations of the holomorphic theory.
In general, a holomorphic boundary condition need not be compatible with such a deformation, but many are; among those that are compatible, some deform to topological boundary theories, whereas others remain holomorphic. Indeed, we will show how to  define a holomorphic boundary condition in the topologically twisted theory, starting from the original $\N=(0,4)$ boundary condition, via a two-step procedure: we first holomorphically twist, and then pass to the topological theory via a further deformation, which is compatible with the holomorphically twisted boundary condition. The deformation of the boundary action required at the level of the untwisted theory is invisible here, since the failure of the boundary condition to be compatible with the deforming supercharge must be exact for the holomorphic supercharge. 

To be concrete, we formulate our boundary conditions in terms of Lagrangian subspaces in the space of boundary fields, not including the case of boundary degrees of freedom that do not originate from bulk fields. Here we follow the formalism of~\cite{ButsonYoo}; their perspective was later developed further by~\cite{BrianOwenEugene}. A deformation of a holomorphic boundary condition then amounts to a deformation of the differential by the deforming supercharge. It is only compatible if the subspace is still a Lagrangian subspace after the deformation. This criterion is both natural and simple from the point of view of the holomorphically twisted theory. We would like to emphasize that the simplification is the merit of the two-step procedure; arguments starting from the initial, physical theory would require a more laborious case-by-case analysis. We show agreement of our results with those of~\cite{CostelloGaiotto} in examples.

We then move on to review Hanany--Witten-style constructions of the relevant 3d $\N=4$ theories. 
To consider boundary conditions, we want to bound the worldvolume of the D3 branes in one further direction. To do so, we introduce further branes, following~\cite{Hanany:2018hlz}. We refer fo these branes as D$5'$ or NS$5'$ branes, where the primes indicate that  they extend in different directions, compared to the initial NS5 and D5 branes. Altogether, the brane settings now preserve $\N=(0,4)$ or $\N=(2,2)$ supersymmetry; holomorphic boundary conditions arise from configurations of $\N=(0,4)$ type. We search for a geometric realization of the deformations studied in~\cite{CostelloGaiotto}, but end by arguing that no obvious geometric deformation of the supports of the branes can result in a family of boundary conditions of an appropriate type (just by studying the related families of Killing spinor equations).  We also briefly review how twisting homomorphisms can be included in string theory constructions, following~\cite{BershadskySadovVafa}; to include the actual operation of taking cohomology of a supercharge, one must consider twisted supergravity.

As  such, we then change perspective, and delve deeper into a geometric construction of twisted worldvolume theories using brane engineering, by  regarding the Hanany--Witten brane constructions in the framework of holomorphically twisted IIB supergravity proposed in~\cite{CostelloLi}.\footnote{S.~Raghavendran informs us that related setups are considered in unpublished work by D.~Butson.} This framework is tailor-made to  mimic the two-step field theory construction on the brane level. 
For backgrounds coupled to D-branes, \cite{CostelloLi} show how further deformations of the supergravity background are inherited by the D-brane world volume theory. To discuss Hanany--Witten brane setups from this perspective, we show how the NS5 and D5 branes we want to include force certain deformations of the IIB background. Including the NS$5'$ and D$5'$ branes that set the boundary conditions on the fields of the 3d theory, we exploit the condition that the required further deformations of the supergravity background need to be compatible with those coming from the unprimed 5-branes. In this way, we derive the deformability of the free hyper and vector multiplet from this novel perspective. Having set the stage for brane engineering in the holomorphic context, we also study the realization of Higgs and Coulomb branches of the gauge theory from Hanany--Witten brane setups.

This paper is organized as follows. In~\S2, we summarize the necessary field theory background. In particular, we review the $\N=4$ algebra and its multiplets and explain the two different topological twists, which we refer to as the $H$- and $C$-twist. We furthermore recall how they can be regarded as deformations of the holomorphic twist. \S3 discusses the deformability of boundary conditions from a field theoretic point of view. To prepare the groundwork, we review the holomorphic twists of the multiplets given in~\S2 as well as boundary conditions of type $(2,2)$ and $(0,4)$. Our main analysis of deformability is given in~\S3.4. \S4 is devoted to review of Hanany--Witten brane setups, and to an analysis of supersymmetries preserved by the configurations coming from rotating these flat-space configurations. 
We find that no obvious geometric deformation can give rise to a family of boundary conditions that is structurally analogous to those constructed by Costello and Gaiotto.
Furthermore, the operation of passing to the cohomology of a supercharge is not clearly visible in the string  theory setup. Both of these points are taken care of in the approach we propose in~\S5, where we use twisted supergravity as formulated in~\cite{CostelloLi}. We start with a brief summary of the relevant main points of their work, and subsequently include Hanany--Witten brane setups into this setting.  Indeed, any brane construction within IIB theory can be studied  from the point of holomorphic brane engineering, which leads to many further applications. We indicate a few of them in our final section.

\subsection*{Acknowledgements}

We thank N.~Paquette, T.~Prochazka, S.~Raghavendran, and B.~Williams for conversations. 
Special thanks are due to K.~Costello, J.~Hilburn, and P.~Yoo for helpful and extensive comments on the draft.
I.S. thanks the University of Heidelberg for hospitality.
I.B. and I.L. are supported by the Deutsche Forschungsgemeinschaft under Germany's Excellence Strategy --- EXC-2094\,--\,390783311 (the ORIGINS Excellence Cluster). I.B. is also supported by the DFG grant ID17448.


\section{Background material}
\label{sec:two}

\subsection{Three-dimensional $\N=4$ supersymmetry and its twists}

\subsubsection{The super-Poincar\'e algebra} 
As with all super-Poincar\'e algebras, the three-dimensional (complex) $\N$-extended super-Poincar\'e algebra is a super Lie algebra which admits a lift to a graded Lie algebra with support in degrees zero, one, and two. The $\Z$-grading is related to the conformal weight.\footnote{In the future, we will make use of a homological grading in discussing physical theories; this grading, by ghost number, is \emph{a priori} unrelated to the integer grading discussed here.}

We let $V = \C^3$ be the abelian Lie algebra of (complexified) translations, which is equipped with a symmetric bilinear pairing, and $\lie{spin}(V) \cong \wedge^2 V$ the Lie algebra of rotations preserving this pairing. Then the Poincar\'e  algebra is the semidirect product 
\[
\lie{aff}(V) = \lie{spin}(V) \ltimes V[-2];
\]
it is the Lie algebra of the group of affine transformations of~$V$.

Let $S$ be the (unique) spinor representation of~$\lie{spin}(V) \cong \lie{su}(2)$, and $R \cong \C^\N$ a complex vector space equipped with a symmetric bilinear pairing $g$. Then the $\N$-extended three-dimensional super-Poincar\'e algebra is 
\deq{
\lie{p_\N}(d=3) = \left( \lie{spin}(V) \oplus \lie{so}(R) \right) \oplus (S \otimes R)[-1] \oplus V[-2].
}
Here, the bracket of degree-one elements is defined to be the tensor product of symmetrization on $S$---using the isomorphism $\Sym^2 S = \lie{sp}(S) = \lie{spin}(V)$---with the symmetric pairing $g$.

We note that the three-dimensional $\N=4$ algebra is obtained by dimensional reduction from the six-dimensional $\N=(1,0)$ super-Poincar\'e algebra. We let $W = \C^6$ be the abelian Lie algebra of six-dimensional translations, $S_+$ the chiral spin representation of~$\lie{spin}(W)$, and $H\cong \C^2$ a two-dimensional vector space equipped with a symplectic pairing. Then the super-Poincar\'e algebra takes the form
\deq{
\lie{p}_{(1,0)}(d=6) = \left( \lie{spin}(W) \oplus \lie{sp}(H) \right) \oplus (S_+ \otimes H)[-1] \oplus W[-2],
}
where the bracket of degree-one elements is defined to be the tensor product of antisymmetrization on~$S_+$ (using the isomorphism $\wedge^2 S_+ \cong W$) with the symplectic form on~$H$. 

To dimensionally reduce this algebra, we choose a splitting $W \cong V \oplus V^\perp$. The stabilizer of this splitting is a subalgebra of the form
\deq{
\stab(V \oplus V^\perp) = \left( \lie{spin}(V) \oplus \lie{spin}(V^\perp) \oplus \lie{sp}(H) \right) \oplus (S_+ \otimes H)[-1] \oplus W[-2] \subseteq \lie{p}.
}
The dimensional reduction is the quotient $\stab(V \oplus V^\perp) / V^\perp$. We can identify this with the three-dimensional $\N=4$ algebra by recalling that, as a $\lie{spin}(V) \oplus \lie{spin}(V^\perp)$-module, $S_+$ decomposes as~$S \otimes C$, where $S$ is the unique spinor of the Lorentz algebra $\lie{spin}(V) \cong \lie{sp}(S)$, as above, and $C$ the spinor of  the other factor $\lie{spin}(V^\perp) \cong \lie{sp}(C)$.
The identification of the $R$-symmetry relies on the accidental isomorphism $\lie{so}(4) \cong \lie{sp}(1) \oplus \lie{sp}(1)$, here in the form
\deq{
\lie{sp}(C) \oplus \lie{sp}(H) \cong \lie{so}(C \otimes H),
} 
where $C \otimes H$ is equipped with a symmetric pairing by tensoring the symplectic pairings on each factor. We thus identify $R \cong C \otimes H$. 

We further remind the reader that a superconformal algebra exists in three dimensions for any value of~$\N$, due to the accidental isomorphism $\lie{spin}(5) \cong \lie{sp}(2)$; the superconformal algebra is then the super Lie algebra $\lie{osp}(\N|2) \subset \lie{gl}(\N|4)$ of linear transformations on an $\N|4$-dimensional supervector space preserving a graded-symmetric bilinear pairing. While this is helpful for remembering the structure of the super-Poincar\'e algebra, it will play no great role for us in what follows.

\subsubsection{Twisting supercharges and twisting homomorphisms}

In the vernacular, ``twisting'' a theory refers to at least two distinct operations. One usage refers to changing (``twisting'') the action of the Lorentz group on the theory on affine space; this requires the choice of a homomorphism $\phi$ from the Lorentz group to the $R$-symmetry group, which we will refer to as the \emph{twisting homomorphism}. On a general spacetime, this means replacing the theory by one in which the fields transform in different bundles---those associated to the twisted Lorentz module structure.

In the context of brane worldvolume theories, where the $R$-symmetry can usually be identified locally with Lorentz rotations transverse to the brane, a choice of twisting homomorphism can be achieved naturally by an appropriate choice of target-space geometry~\cite{BershadskySadovVafa}. On the other hand, taking $Q$-cohomology is not visible in any obvious geometric way in a string theory construction; rather, it corresponds to working in a background where the corresponding ghost acquires a vacuum expectation value, as in the formulation of twisted supergravity by Costello and Li~\cite{CostelloLi}. We return to these points in more detail in what follows. 

Having chosen a twisting homomorphism, it may be that some (not necessarily unique) supercharge $Q$ is invariant under the twisted Lorentz symmetry.
Since there is no scalar translation, it follows just for representation-theoretic reasons that $[Q,Q]=0$; moreover, the scalar supercharge acts on the theory even after it is placed on a general smooth manifold. The other meaning of ``twisting'' then refers to the procedure of taking the invariants of this scalar supercharge, or (equivalently) of deforming the BRST or BV differential of the theory by~$Q$. The result is a topological field theory defined on smooth manifolds, for which the action of translations is nullhomotopic with respect to~$Q$.

However, if we do not insist that the resulting theory is defined on all smooth manifolds, more general kinds of twists exist. To think of them, it is useful to imagine reversing the order of the logic sketched above. The super-Poincar\'e group acts on the theory on flat space, and any supercharge of square zero defines a possible twist. We can then ask about the stabilizer of~$Q$ as a subgroup of the original Lorentz and $R$-symmetry; if this contains a subgroup isomorphic to the full Lorentz group, it reflects the existence of a twisting homomorphism of the type discussed above. If not, though, we can still define the $Q$-twisted theory on manifolds whose holonomy group maps into the stabilizer of~$Q$ (in such a way that the translations transform appropriately).

It is thus important to classify the supercharges of square zero (or equivalently Maurer--Cartan elements) in the super-Poincar\'e algebra, as was done systematically in~\cite{NV,ChrisPavel}. We review this classification for three-dimensional $\N=4$ supersymmetry here. Odd elements of $\lie{p}_4(d=3)$ lies in the tensor product $S \otimes C \otimes H$; the bracket is the tensor product of symmetrization on~$S$ with antisymmetrization (symplectic contraction) on~$C$ and on~$H$. 

There are three classes of supercharges of square zero:
\begin{enumerate}
\item A decomposable element $Q = \psi \otimes c \otimes h \in S \otimes C \otimes H$. The image of~$Q$ is a two-dimensional subspace of~$V$. Such an element is a minimal or holomorphic supercharge. 
\item An element of the form $(\psi_1 \otimes c_1 + \psi_2 \otimes c_2) \otimes h$, which is the tensor product of an element $h$ with a full-rank matrix in $S \otimes C$. The image of~$Q$ is all of~$V$, and the supercharge is thus topological. Particular supercharges of this type are picked out by the twisting homomorphism $\phi_C: \lie{sp}(S) \rightarrow \lie{sp}(C)$. These supercharges are used in the Rozansky--Witten twist.
\item An element, constructed in similar fashion, which is the tensor product of an element $c\in C$ with a full-rank matrix in $S\otimes H$. These are again topological; a compatible twisting homomorphism is $\phi_H: \lie{sp}(S) \rightarrow \lie{sp}(H)$.
\end{enumerate}

Geometrically, $S \otimes C \otimes H$ can be viewed as a space of two-by-four matrices in three different ways; the topological supercharges identified above are those that are of rank one with respect to two of these decompositions, either as $(S \otimes C)\otimes H$ or as $(S \otimes H) \otimes C$. As such, the spaces of such supercharges are two copies of Segre-embedded $P^1 \times P^3$, intersecting along $P^1 \times P^1 \times P^1$ (precisely at the locus of holomorphic supercharges). See~\cite{NV} for more details.

These considerations make it clear that any topological twist can be constructed as a deformation of a holomorphic twist. Later in the paper, we will fix a chosen holomorphic supercharge and study the corresponding holomorphically twisted multiplets. The perturbing supercharges that deform this holomorphic twist to topological twists of $C$ or $H$ type will act on the holomorphically twisted theory. 

We note that there is an obvious ``mirror'' outer automorphism of $\lie{p}_4(d=3)$, which acts by reversing the tensor factors $C$ and~$H$, and correspondingly exchanges the two classes of topological supercharges above. There is also an action of the mirror map on the set of multiplets, by pulling back along the mirror automorphism of the algebra; pairs of multiplets 
can thus be related by this $\Z/2\Z$ action, to which we turn in more detail below. To distinguish between $C$ and~$H$ in an invariant manner, it is necessary to think about lifting either to four or to six dimensions; as reviewed above, $\lie{sp}(H)$ can be identified with the six-dimensional $R$-symmetry, whereas $\lie{sp}(C)\cong\lie{spin}(V^\perp)$ appears upon dimensional reduction.

\subsubsection{Multiplets}

We briefly remind the reader of the multiplets we will use in what follows. The vector multiplet and hypermultiplet in three-dimensional $\N=4$ supersymmetry are easiest to construct by dimensional reduction from six dimensions; conjugation by the mirror automorphism
exchanges the hypermultiplet with a conjugate multiplet, normally called the ``twisted hypermultiplet'' in the literature. To avoid overburdening the word ``twist'' more than is necessary, we will refer to these as $H$- and $C$-hypermultiplets respectively, according to the representation in which the scalar fields transform. Thanks to a pleasing notational coincidence, $H$ should remind the reader of ``hypermultiplet,'' and $C$ of ``conjugate hypermultiplet.''

The physical fields of the six-dimensional $\N=(1,0)$ vector multiplet are an even gauge field $A \in \Omega^1(\R^6) \otimes \lie{g}$ and a fermion $\psi \in \Pi S_+(\R^6) \otimes H \otimes \lie{g}$. Here $\lie{g}$ is the semisimple Lie algebra associated to the gauge group.

Upon dimensional reduction, $S_+$ becomes the tensor product of the three-dimensional spin bundle $S(\R^3)$ with $C$, as discussed above, and $\Omega^1$ becomes the direct sum of one-forms on the three-manifold with scalars valued in $\Sym^2(C) \cong \lie{sp}(C)$. So the physical field content of the three-dimensional $\N=4$ vector multiplet consists of
\deq{
A \in \Omega^1(\R^3) \otimes \lie{g}, \qquad
\phi \in \Omega^0(\R^3) \otimes \Sym^2(C) \otimes \lie{g}, \qquad
\psi \in \Pi S(\R^3) \otimes C \otimes H \otimes \lie{g}.
}

We now turn to the six-dimensional $\N=(1,0)$ hypermultiplet. This multiplet takes values in a symplectic vector space, which we will call $(F,\omega)$ to emphasize that $\lie{sp}(F)$ is a  flavor symmetry. Its physical fields consist of scalars $\phi \in \Omega^0(\R^6) \otimes H \otimes F$,  as well as fermions that are neutral under the $R$-symmetry, $\psi \in \Pi S_+(\R^6) \otimes F$.

Upon dimensional reduction, we find that the physical field content of the three-dimensional $\N=4$ ($H$-)hypermultiplet consists of
\deq{
\phi \in \Omega^0(\R^3) \otimes H \otimes F, \qquad
\psi \in \Pi S(\R^3) \otimes C \otimes F.
}
Of course, the physical fields of the conjugate hypermultiplet are the same, but with the roles of $C$ and $H$ reversed.

After a modification to the action of  the~$\lie{sp}(C)$ $R$-symmetry, the vector multiplet and the conjugate hypermultiplet are  related by electric-magnetic duality. The physical field content of the  conjugate hypermultiplet is
\deq{
\phi \in \Omega^0(\R^3) \otimes C \otimes F, \qquad
\psi \in \Pi S(\R^3) \otimes H \otimes F.
}
Since the flavor symmetry  does  not  act on the  supersymmetry  algebra  itself,  we can modify the action of  $R$-symmetry on  this multiplet  in a fashion somewhat analogous to the  twisting homomorphisms  discussed above. In  particular, we can just as well let the $R$-symmetry  $\lie{sp}(C)$  act via the  diagonal embedding in $\lie{sp}(C) \oplus \lie{sp}(F)$. After doing this, the field content of the $C$-hypermultiplet becomes 
\deq{
\phi \in \Omega^0(\R^3) \otimes C^{\otimes 2}, \qquad
\psi \in \Pi S(\R^3) \otimes H \otimes C.
}
We recall that $C^{\otimes 2} = \Sym^2(C)  \oplus \C$,  so that one  scalar field is  now  neutral under $R$-symmetry. Electric-magnetic duality replaces this scalar field by a  one-form,  so that  we recover the  field content of the  abelian vector multiplet. A similar story relates the image of the vector multiplet under the mirror automorphism to the $H$-hypermultiplet. The ambiguity in the action of the $R$-symmetry on multiplets discussed here will recur in the sequel, where different choices arise naturally from different background geometries in the context of $D$-brane worldvolume theories.

We do not recall the supersymmetry transformations or the formulation of these multiplets as BV theories explicitly here; the reader is referred to the literature~\cite{SW3d,IntriligatorSeiberg}. For our purposes, it is sufficient to recall the form of the holomorphically twisted multiplets in the BV formalism, which we do below.
For the purposes of discussing twisting homomorphisms, though, it is useful to recall how the fields of each multiplet transform after applying the twisting homomorphisms. Recall that the effect of the twisting homomorphism $\phi_C$ (resp.~$\phi_H$) is to replace $C$ (resp.~$H$) by a copy of the spin bundle. (In fact, one typically uses the twisting homomorphism together with a regrading datum, so that $C$ becomes a copy of the odd spin bundle, shifted appropriately in homological degree; we will not review the details of this here.) We quickly summarize the results of this procedure in Table~\ref{tab: twisted}.
\begin{table}
\begin{equation*}
\begin{array}{l|c|l}
\text{hypermultiplet} & {\phi_C} & \phi \in \Omega^0(\R^3) \otimes H \otimes F \\
&  &  \psi \in \Omega^1(\R^3) \otimes F[-1] \oplus \Omega^0(\R^3)\otimes F[-1]  \\ \hline
\text{hypermultiplet} & \phi_H & \phi \in \Pi S(\R^3)\otimes F[-1] \\ 
& & \psi \in \Pi S(\R^3) \otimes C \otimes F \\ \hline \hline
\text{vector multiplet} & \phi_C & A \in \Omega^1(\R^3) \otimes \lie{g} \\
& & \psi \in \psi \in \Omega^1(\R^3) \otimes H \otimes \lie{g} [-1] \oplus \Omega^0(\R^3)\otimes H \otimes \lie{g} [-1]  \\
& & \phi \in \Omega^1(\R^3)\otimes\lie{g}[-2] \\ \hline
\text{vector multiplet} & \phi_H & A \in \Omega^1(\R^3) \otimes \lie{g} \\
& & \psi \in \psi \in \Omega^1(\R^3) \otimes C \otimes \lie{g} [-1] \oplus \Omega^0(\R^3)\otimes C \otimes \lie{g} [-1]  \\
& & \phi \in \Omega^0(\R^3)\otimes\Sym^2(\C) \otimes \lie{g} 
\end{array}
\end{equation*}
\caption{Field content of 3d $\N=4$ multiplets after applying twisting homomorphisms}
\label{tab: twisted}
\end{table}

\subsection{A brief reminder on the BV approach}

In this section, we give a few schematic words of reminder about the Batalin--Vilkovisky formalism for quantum field theories, which we will use in discussing the field-theoretic perspective on deformable boundary conditions below. We will not consider quantization in any detail; for that reason, our discussion focuses on classical aspects of BV theories. 

A classical field theory can be thought of as a particular sheaf, which assigns to an open subset $U$ of the spacetime the space of solutions of the equations of motion over~$U$ up to gauge equivalence. The equations of motion arise from a variational problem, associated to an action functional $S$ on sections of a (finite-rank differential graded affine) vector bundle over the spacetime. 

While the sheaf of off-shell fields is locally free, the sheaf of on-shell field configurations is not; in particular, since we are in general interested in well-defined boundary value problems, a section of the sheaf over~$U$ should be uniquely determined by its boundary values along~$\partial U$. We can identify the space of sections over~$U$ with the phase space of the theory on~$\partial U$, which should in general carry a symplectic structure. 

The BV formalism replaces the sheaf of solutions to the equations of motion by the \emph{derived critical locus}, which carries a canonical $(-1)$-shifted symplectic structure~\cite{PTVV}. Just as the BRST formalism accomplishes the quotient by gauge symmetries homologically by adding ghost fields of homological degree $-1$ corresponding to gauge symmetry generators, the BV formalism adds ``antifields'' of degree  $+1$ that correspond to the equations of motion, together with  a differential that imposes those equations on the physical fields in degree zero. Since the equations of motion arise variationally, they correspond naturally to the fields, giving rise to a pairing between fields and antifields of homological degree $-1$ that is the shifted symplectic structure mentioned above. Noether's second theorem ensures that gauge symmetries correspond one-to-one to syzygies between the equations of motion, requiring ``antighosts'' of degree $+2$ that pair appropriately with the ghosts and ensuring that the shifted symplectic structure is also present in gauge theories. 

Just as in classical symplectic geometry, the algebra of functions on a $(-1)$-shifted symplectic space acquires a $(+1)$-shifted Poisson bracket, often called the ``antibracket'' in the BV literature. This bracket sets up a correspondence between degree-zero functions and degree-$(+1)$ derivations that are the corresponding Hamiltonian vector fields; the derivation associated to a functional $S$ squares to zero when the ``classical master equation'' $\{S,S\} = 0$ is satisfied. This is automatic for the action  functional of the original theory, since it is independent of the antifields; the differential $\{S,-\}$ is then precisely the one mentioned above that imposes  the variational equations of motion on the physical fields. If present, the BRST differential can also be extended to a Hamiltonian vector field and thus encoded in further terms in the BV action functional.

Quantization can then be accomplished by a further deformation of the differential, by the ``BV Laplacian'' (a canonical second-order differential operator). See~\cite{Owen-thesis} for a more detailed discussion of the formalism, from a perspective similar to ours, and~\cite{Wthesis} for further discussion in the context of holomorphic field theories, relevant to the examples of holomorphic twists we consider below. The reader who is interested in the original literature should examine \cite{BV,Schwarz-BV,Severa}, just for example.


\section{The field theory perspective on deformable boundary conditions}
\label{sec: FT}

\subsection{Holomorphic twists of three-dimensional $\N=4$ theories}
\subsubsection{The holomorphic twist of the hypermultiplet}
The holomorphic twist of the three-dimensional $\N=4$ hypermultiplet is known in the literature~\cite{TCSMT,CostelloDimofteGaiotto,ESW}, but we recall it quickly here. It is easiest to write it down by considering a dimensional reduction of the six-dimensional $\N=(1,0)$ hypermultiplet. 

Assuming a flat target, the hypermultiplet takes values in a symplectic vector space, as recalled above. We again denote this space $(F,\omega)$. In six dimensions, the corresponding holomorphic theory is a fermionic analogue of holomorphic abelian Chern--Simons theory. Its BV fields are given by
\deq{
\E^\bullet = \Omega^{0,*}(\C^3)[1] \otimes K^{1/2}_{\C^3} \otimes \Pi F.
}
(Abelian Chern--Simons theory can be defined with values in a super-vector space equipped with graded symmetric pairing; the relevant example for us is the parity reversal of a symplectic vector space.)

To compute the dimensional reduction, we note that reducing along a single real direction has the effect of replacing the corresponding factor $\Omega^{0,*}(\C)$ in the Dolbeault complex by the de Rham complex $\Omega^*(\R)$, one dimension lower. Reducing along the other real direction leaves only a copy of the cohomology of the circle, which we write as $\C[\epsilon]$ and implicitly equip with the pairing given by the Berezin integral. (Here $\epsilon$ is a formal parameter of homological degree one, which can be thought of as the remaining fermionic zero mode $d\bar{z}$.) 

As such, the dimensional reduction of a single copy of the Dolbeault complex on~$\C^3$ (which we think of as freely resolving the sheaf of holomorphic functions there) is  the differential graded algebra
\deq{
\O[\epsilon] =  \Omega^{0,*}(\C) \otimes_\C \Omega^*(\R) \otimes_\C \C[\epsilon].
}
We use the shorthand $\O$ for the free resolution of the sheaf functions that are holomorphic in two of the three spacetime directions and locally constant in the third, by the tensor product of the Dolbeault and de Rham complexes. $\O$ has a pairing of degree $-2$ with $\O \otimes K_\C$.

The dimensionally reduced theory is thus isomorphic to 
\deq{
\E^\bullet = \O[1] \otimes K^{1/2}_{\C} \otimes \Pi F[\epsilon].
}
(The other square roots of canonical bundles can be absorbed into the choice of grading on $\C[\epsilon]$.)
The reader will notice that the theory is of rank four over $\O$, containing two bosonic and two fermionic holomorphic-topological fields. 

Moreover, it is important to notice that the physical scalar fields in the hypermultiplet sit in homological degree $-1$ after the twist. (This occurs because the six-dimensional $R$-symmetry acts nontrivially on the \emph{scalars}, rather than the fermions, in the hypermultiplet.)

\subsubsection{The holomorphic twist of the vector multiplet}
The holomorphic twist of the vector multiplet is again well-known in the literature. We again recall it using dimensional reduction from six dimensions, where the twist of the (perturbative) $\N=(1,0)$ vector multiplet with gauge algebra $\lie{g}$ is the holomorphic BF theory whose fields are
\deq{
\E^\bullet = \Omega^{0,*}(\C^3)[1] \otimes \lie{g}  \oplus \Omega^{3.*}(\C^3)[1] \otimes \lie{g}^\vee.
}
The reduction to three dimensions proceeds in similar fashion; the result is
\deq{
\E^\bullet = \O[1] \otimes \lie{g}[\epsilon]  \oplus \left( \O\otimes \K_\C \right)  \otimes (\lie{g}[\epsilon])^\vee,
}
where we again use the shorthand $\O = \Omega^{0,*}(\C) \otimes_\C \Omega^*(\R)$ for the free resolution of the sheaf of functions that are holomorphic in two of the three directions and locally constant in the third. Note that we also used the fact that $\lie{g}^\vee[\epsilon] = (\lie{g}[\epsilon])^\vee[-1]$; this notation makes it easier to see the decomposition into the holomorphic twist of the $\N=2$ vector and chiral multiplets. (The $\N=2$ vector is associated to the term $\lie{g}$ and the adjoint chiral to $\epsilon \lie{g}$.)

We emphasize that the physical vector field, as well as the scalar field in the $\N=2$ adjoint chiral multiplet, sit in homological degree zero in the twisted theory. 

\begin{figure}[t]
\[
\begin{tikzcd}[row sep = 2 pt]
-1 & 0 & 1 & 2 \\ \hline \\
\lie{g} \otimes \Omega^{0,0;0}\ar[r,"\partial"] \ar[dr,"d"']& \lie{g} \otimes \Omega^{0,1;0} \ar[dr,"d"] \\
&\lie{g} \otimes  \Omega^{0,0;1}\ar[r,"\partial"'] &\lie{g} \otimes  \Omega^{0,1;1} \\[1ex]
& \epsilon\lie{g} \otimes  \Omega^{0,0;0} \ar[r,"\partial"] \ar[dr,"d"']& \epsilon\lie{g} \otimes  \Omega^{0,1;0} \ar[dr,"d"] \\
& & \epsilon\lie{g} \otimes  \Omega^{0,0;1} \ar[r,"\partial"']&\epsilon\lie{g} \otimes  \Omega^{0,1;1} \\[1ex]
(\epsilon \lie{g})^\vee \otimes \Omega^{1,0;0} \ar[r,"\partial"] \ar[dr,"d"']& (\epsilon \lie{g})^\vee \otimes \Omega^{1,1;0} \ar[dr,"d"] \\
&(\epsilon \lie{g})^\vee \otimes  \Omega^{1,0;1} \ar[r,"\partial"']& (\epsilon \lie{g})^\vee \otimes \Omega^{1,1;1} \\[1ex]
& \lie{g}^\vee \otimes \Omega^{1,0;0} \ar[r,"\partial"] \ar[dr,"d"']&\lie{g}^\vee \otimes \Omega^{1,1;0} \ar[dr,"d"] \\
& & \lie{g}^\vee \otimes \Omega^{1,0;1} \ar[r,"\partial"']&\lie{g}^\vee \otimes \Omega^{1,1;1} \\ \hline \\
\Pi F \otimes \Omega^{0,0;0}\ar[r,"\partial"] \ar[dr,"d"']& \Pi F \otimes \Omega^{0,1;0} \ar[dr,"d"] \\
&\Pi F \otimes  \Omega^{0,0;1}\ar[r,"\partial"'] &\Pi F \otimes  \Omega^{0,1;1} \\[1ex]
& \epsilon\Pi F \otimes  \Omega^{0,0;0} \ar[r,"\partial"] \ar[dr,"d"']& \epsilon\Pi F \otimes  \Omega^{0,1;0} \ar[dr,"d"] \\
& & \epsilon\Pi F \otimes  \Omega^{0,0;1} \ar[r,"\partial"']&\epsilon\Pi F \otimes  \Omega^{0,1;1} 
\end{tikzcd}
\]
\caption{Holomorphic twists of three-dimensional $\N=4$ multiplets}
\end{figure}

\subsection{Boundary conditions in holomorphic theories}

To think about simple classes of boundary conditions in the BV formalism, we will use the framework of~\cite{ButsonYoo}, which is adapted to field theories that are \emph{topological normal to the boundary}. (For additional background and more general approaches to BV theories with boundary, see~\cite{CMR,BrianOwenEugene,Rabinovich}.) Concretely, this means that the equations of motion require fields to be locally constant in the normal direction; the space of BV fields on a half-space thus factorizes as
\deq{
\E^\bu = \E^\bu_\partial \otimes \Omega^*(\R_+),
}
where the \emph{space of boundary fields} $\E^\bu_\partial$, just like the normal phase space, is 0-shifted symplectic. A boundary condition is then presented by giving a (graded) Lagrangian subspace $\L \subset \E^\bu_\partial$ of the space of boundary fields. For clarity, we show the spaces $\E^\bu_\partial$ of boundary fields for the minimally twisted vector multiplet and hypermultiplet in Figure~\ref{fig:bdy}.

\begin{figure}[t]
\[
\begin{tikzcd}[row sep = 2 pt]
-1 & 0 & 1  \\ \hline \\
\lie{g} \otimes \Omega^{0,0}\ar[r,"\partial"] & \lie{g} \otimes \Omega^{0,1}\\
& \epsilon\lie{g} \otimes  \Omega^{0,0} \ar[r,"\partial"] & \epsilon\lie{g} \otimes  \Omega^{0,1}  \\
(\epsilon \lie{g})^\vee \otimes \Omega^{1,0} \ar[r,"\partial"] & (\epsilon \lie{g})^\vee \otimes \Omega^{1,1} \\
& \lie{g}^\vee \otimes \Omega^{1,0} \ar[r,"\partial"] &\lie{g}^\vee \otimes \Omega^{1,1} \\ \hline \\
\Pi F \otimes \Omega^{0,0}\ar[r,"\partial"] & \Pi F \otimes \Omega^{0,1}\\
& \epsilon\Pi F \otimes  \Omega^{0,0} \ar[r,"\partial"] & \epsilon\Pi F \otimes  \Omega^{0,1}
\end{tikzcd}
\]
\caption{Boundary fields of twisted three-dimensional $\N=4$ multiplets}
\label{fig:bdy}
\end{figure}

\subsubsection{Boundary conditions of $\N=(0,4)$ type}

\paragraph{The hypermultiplet}
We recall that, as studied in~\cite{CostelloGaiotto}, there are two evident ways to impose half-BPS boundary conditions on the three-dimensional $\N=4$ hypermultiplet that preserve a two-dimensional $\N=(0,4)$ subalgebra. We may impose either Neumann or Dirichlet boundary conditions uniformly on all of the hypermultiplet scalars. Corresponding boundary conditions on fermions are then determined by supersymmetry.
Such boundary conditions are obviously compatible with the holomorphic twist.

In our formalism, a boundary condition for the holomorphically twisted theory consists of a graded Lagrangian submanifold of the space $\E^\bu_\partial$ of boundary fields. A Dirichlet boundary condition corresponds to setting a particular field to zero, whereas a Neumann boundary condition corresponds to imposing no condition on the field (and, in fact, setting its corresponding antifield to zero). We have recalled above that the boundary values of those hypermultiplet scalars that survive the holomorphic twist sit in $\Pi F \otimes  \Omega^{0,0}  [1]$, the leftmost term in Figure~\ref{fig:bdy}.

It is thus immediate to understand the Lagrangian submanifolds corresponding to the holomorphic twists of the above $\N = (0,4)$ boundary conditions: we have
\begin{equation}
\begin{aligned}
\L_N = \Pi F \otimes \Omega^{0,*}[1] &\subset \Pi F[\epsilon] \otimes \Omega^{0,*}[1] = \E^\bu_\partial; \\
\L_D = \epsilon \Pi F \otimes \Omega^{0,*}[1] &\subset \Pi F[\epsilon] \otimes \Omega^{0,*}[1] = \E^\bu_\partial.
\end{aligned}
\end{equation}

\paragraph{The vector multiplet}
The simplest way to understand the analogous $\N=(0,4)$ boundary conditions for the vector multiplet is to recall the equivalence with the $C$-hypermultiplet under electric-magnetic duality discussed above. Since electric-magnetic duality should reverse Dirichlet and Neumann boundary conditions, we expect that Neumann boundary conditions for the vector multiplet scalars should be paired with Dirichlet boundary conditions for the gauge field, and vice versa. As reviewed in~\cite{CostelloGaiotto}, this turns out to be correct. We will use the terms ``Neumann'' and ``Dirichlet'' to refer to the boundary conditions on scalars, since this seems to be the least confusing choice. 

Looking at the space of boundary fields in~Figure~\ref{fig:bdy}, we recall that the physical gauge field gives rise to the degree-zero piece of $\lie{g} \otimes \Omega^{0,*}[1]$, whereas the surviving physical scalars live in the degree-zero piece of $\epsilon \lie{g} \otimes \Omega^{0,*}[1]$. This makes it easy to read off the Lagrangian submanifolds corresponding to the holomorphic twists of boundary conditions of $\N=(0,4)$ type:
\begin{equation}
\begin{aligned}
\L_N = \epsilon \lie{g} \otimes \Omega^{0,*}[1] \oplus \lie{g}^\vee \otimes \Omega^{1,*} &\subset \lie{g}[\epsilon] \otimes \Omega^{0,*}[1]  \oplus (\lie{g}[\epsilon])^\vee\otimes \Omega^{1,*} = \E^\bu_\partial; \\
\L_D = \lie{g} \otimes \Omega^{0,*}[1] \oplus (\epsilon \lie{g})^\vee \otimes \Omega^{1,*} &\subset \lie{g}[\epsilon] \otimes \Omega^{0,*}[1]  \oplus (\lie{g}[\epsilon])^\vee\otimes \Omega^{1,*} = \E^\bu_\partial.
\end{aligned}
\end{equation}

\subsubsection{Boundary conditions of $\N=(2,2)$ type}

Although we are primarily interested in (deformations of) the boundary conditions of $\N=(0,4)$ type reviewed above, we briefly remind the reader of how standard boundary conditions of $\N=(2,2)$ type look in our language. Such boundary conditions are automatically compatible with both the $C$ and $H$ twists. 

For the hypermultiplet, we can construct an $\N=(2,2)$ boundary condition by, roughly speaking, imposing Dirichlet boundary conditions on half of the scalar fields and Neumann boundary conditions on the other half. More accurately, we choose a Lagrangian submanifold $L \subset F$ of the flavor space, and decompose $F = L \oplus L^\vee$. Then we impose Neumann boundary conditions on the scalars in~$L^\vee$ and Dirichlet boundary conditions on those in~$L$. Understanding the associated Lagrangian submanifold in the space of holomorphic boundary fields is trivial; it is just 
\deq{
\L_{(2,2),L} = \Pi L[\epsilon] \otimes \Omega^{0,*}[1] \subset \Pi F[\epsilon] \otimes \Omega^{0,*}[1] = \E^\bu_\partial.
}

For the vector multiplet, boundary conditions of $\N=(2,2)$ type will be reviewed below in~\S\ref{sec:Three}.
In our framework, it is easy to write the corresponding Lagrangian submanifolds in the space of boundary fields: they are just
\begin{equation}
\begin{aligned}
\L_{(2,2),N} = \lie{g}[\epsilon] \otimes \Omega^{0,*}[1] &\subset \lie{g}[\epsilon] \otimes \Omega^{0,*}[1]  \oplus (\lie{g}[\epsilon])^\vee\otimes \Omega^{1,*} = \E^\bu_\partial; \\
\L_{(2,2),D} = (\lie{g}[\epsilon])^\vee \otimes \Omega^{1,*} &\subset \lie{g}[\epsilon] \otimes \Omega^{0,*}[1]  \oplus (\lie{g}[\epsilon])^\vee\otimes \Omega^{1,*} = \E^\bu_\partial.
\end{aligned}
\end{equation}

\subsection{Topological twists as deformations of the holomorphic theory}

We now wish to identify the action of the supercharges which deform the holomorphically twisted theory to the $C$ or $H$ topological twists. Properly speaking, the holomorphically twisted theory admits an action of the dg Lie algebra 
\deq{
\lie{p}_Q =  (\lie{p},  \ad_Q),
}
obtained by deforming the super-Poincar\'e algebra by the holomorphic supercharge (viewed as a  Maurer--Cartan element). We will usually work at the level of the cohomology $H^*(\lie{p}_Q)$. 

Let us understand the structure of this dg Lie algebra explicitly. As reviewed  above,  a holomorphic supercharge is a choice of decomposable element  
\deq{
Q =  s \otimes c \otimes h \in S \otimes C  \otimes H;
}
it therefore determines a polarization  of  each of those three two-dimensional symplectic vector spaces.  We fix a corresponding Darboux basis  of each. The  structure of the dg Lie algebra is shown in Figure~\ref{fig:pQ}.

\begin{figure}
\begin{equation*}
\begin{tikzcd}[row sep = 3 pt]
0  &  1 &  2 \\ \hline
\lie{gl}(s) \ar[dr] \\
\lie{gl}(c) \ar[r] &  s \otimes c \otimes h  &  (s^\vee)^{\otimes 2}  \\
\lie{gl}(h)\ar[ur] \\ 
(s^\vee)^{\otimes 2} \ar[r] &  s^\vee \otimes c \otimes h \\
(c^\vee)^{\otimes 2} \ar[r] &  s \otimes c^\vee \otimes h \\
(h^\vee)^{\otimes 2} \ar[r] &  s \otimes c \otimes h^\vee \\
s^{\otimes 2} &  s \otimes c^\vee \otimes h^\vee \ar[r] &  s^{\otimes 2} \\
c^{\otimes 2} &  s^\vee \otimes c^\vee \otimes h^\vee \ar[r]  &  \lie{gl}(s) \\
h^{\otimes 2} & s^\vee \otimes c \otimes h^\vee & \\
 & s^\vee \otimes c^\vee \otimes h &
\end{tikzcd}
\end{equation*}
\caption{The dg Lie algebra $\lie{p}_Q$ for  holomorphically  twisted $\N=4$  theories in three dimensions}
\label{fig:pQ}
\end{figure}

Using this diagram, it  is  straightforward to  understand the cohomology $H^*(\lie{p}_Q)$. There is  one surviving holomorphic translation $P$ in degree two, denoted $(s^\vee)^{\otimes 2}$. Two odd elements survive; these are the deforming supercharges to  each of the topological  twists, $Q_C = s^\vee \otimes c^\vee \otimes h$ and $Q_H = s^\vee \otimes c \otimes h^\vee$. The bracket is just $[Q_C, Q_H] = P$. Finally, the  degree zero cohomology  consists of the  parabolic Lie algebra that stabilizes  the holomorphic supercharge.

It is also straightforward to understand how such supercharges act on the holomorphically twisted theories discussed above. We know  that holomorphic translations must act via $P = \partial/\partial z$, so it remains to find two odd operators with the correct anticommutator. This is easiest to do by remembering the dimensional reduction computation above. The $C$-twist originates from a generic dimensional reduction of  the six-dimensional holomorphic twist, so that $Q_C$ must look like a Beltrami differential in the unreduced theory. (To see this, remember that a nilpotent supercharge in six dimensions is a rank-one element of $S_+ \otimes H$, which however can be of arbitrary rank with respect to the decomposition $S_+ \cong S \otimes C$.) Recalling that $\epsilon$ originated from $d\zbar_3$, it is easy to see that 
\deq{
Q_C = \epsilon \pdv{ }{z}, \quad Q_H = \pdv{ }{\epsilon}.
}
($Q_H$ is then the unique odd endomorphism with the correct commutation relations.) To understand the action of this algebra on the vector multiplet, we recall that the Berezin integral identifies $(\lie{g}[\epsilon])^\vee \cong \lie{g}^\vee[\epsilon][1]$.

\subsection{Deformability of holomorphic boundary conditions}

In light of the discussion above, it is easy to formulate a notion of compatibility for a boundary condition $i: \L \to \E_\partial$. We let $\E_\partial^C$ and $\E_\partial^H$ denote the deformations of the cochain complex $\E_\partial$ by the respective supercharges $Q_C$ and $Q_H$ constructed above. 
\begin{dfn}
Fix a boundary condition $\L \subset \E_\bdy$ and a deformation $Q$ of~$\E^\bdy$, which we think of as a deformation of the differential. Then $\L$ is compatible with the chosen deformation (or ``deformable'') precisely when 
\deq{
i : \L \to (\E_\partial,d_\E + Q)
} is a cochain map when viewed as a map to the deformed space of boundary fields. In other words,  the subspace $\L$ is preserved by  the deformation of the differential, $Q(i(\L)) \subset i(\L)$. 
\end{dfn}

We remark on the interpretation of this definition. Of course, a boundary condition of the deformed theory is, in our sense, nothing other than a Lagrangian subspace of the deformation of the space of boundary fields. The definition says that $\L$ is such a subspace, \emph{without} any deformation: in other words, that the inclusion map $i$ exhibits $\L$ as  a Lagrangian subobject of~$\E_\bdy$ for any value of the deformation parameter whatsoever. From the perspective of the holomorphically twisted theory, $\L$ is just compatible with the deformation on the nose. (Of course, the differential on~$\L$ may be deformed by~$Q$, as happens in the boundary conditions of $\N=(2,2)$ type. With respect to the decomposition $\E_\bdy = \L \oplus \L^\vee$, the only term forbidden by the definition is a map from $\L$ to~$\L^\vee$. Maps from~$\L^\vee$ to~$\L$ can and do occur, as in the examples of $\N=(0,4)$ type. Endomorphisms of~$\L$ occur in the examples of~$\N=(2,2)$ type; the corresponding endomorphism of~$\L^\vee$ is then determined by the condition that the differential be an odd symplectic  vector field.)

Why, then, do we term such boundary conditions ``deformable''? 
To make this precise, we restrict for simplicity to the situation when the undeformed boundary fields $\E_\bdy$ are just isomorphic to $T^*\L = \L\oplus \L^\vee$ and the deforming supercharge does not induce any new differential on~$\L$.
Let $\hat{d}$ denote the combination of the BV differential and the holomorphic supercharge that acts on the untwisted theory, and~$d$ the internal differential on~$\L$.
Let us choose a deformation retraction of $(\hat \L, \hat d)$ onto~$(\L,d)$, consisting (in standard fashion) of a projection, an inclusion, and a nullhomotopy of the composite endomorphism $jp$ of~$(\hat \L, \hat d)$:
\begin{equation}
\begin{tikzcd}
\arrow[loop left]{l}{h}  (\hat \L, \hat d)  \ar[r, two heads, shift left, "p"] &
(\L,d).  \ar[l, right hook->, shift left, "j"] 
\end{tikzcd}
\end{equation}
 In the situation above, this obviously determines a homotopy datum relating $\hat\E_\bdy$ to~$\E_\bdy$. 

In the holomorphic theory, $\L$ is compatible with the deforming supercharge $Q$; in particular, our definition insists that the component $Q_{01}$ of $Q$ that maps $\L$ to~$\L^\vee$ is zero. However, this does not mean that a boundary condition $\hat\L$ of the
full theory need be compatible, in this strict sense, with \emph{any} supercharge $\hat Q$ that induces the deformation of the holomorphic theory! Indeed, all we can know is that 
$\hat Q_{01} \hat \L \subset \hat \L^\vee$ is exact with respect to~$\hat{d}$.

An application of the homological perturbation lemma then defines a new subspace $\hat \L_\text{def}$ of the untwisted boundary fields, which is,  at least morally speaking, just the graph of the $\hat{d}$-nullhomotopy of $\hat Q_{01} \hat \L$. Concretely, we can consider the degree-zero cochain map
\[
\varphi = h \hat{Q}_{01}: \hat \L \to \hat \L^\vee
\]
that witnesses the $\hat d$-exactness of~$\hat Q_{01}$. 
Then one can consider the graph of this map:
\[
{i}' = (i, \varphi): \hat\L \to \left( \hat\L \oplus \hat\L^\vee, \hat d + \hat Q \right).
\]
If $\hat Q_{10}$ then annihilates $\varphi$, we are done, and the deformed boundary condition is literally the graph of~$\varphi$; otherwise, we will need to correct $i$ to compensate for the $\hat Q_{10}$ variation of~$\varphi$, and a chain of  further corrections may be necessary (which, however, will remain finite for degree reasons). Since it is not our main focus to study the deformations at the level of the  untwisted theory, we do not pursue this further, and return to the setting of the holomorphic twist.

Having set up our terms, it is then a matter of inspection to recover the following result:
\begin{fact}[\cite{CostelloGaiotto}] 
Consider the holomorphically twisted 3d $\N=4$ hypermultiplet, with space of boundary fields $\E_\partial = \Omega^{0,*}(\C)[1] \otimes \Pi F[\epsilon]$. Then:
\begin{itemize}[itemsep = 0 pt]
\item the boundary condition $\L_N$ is compatible with $Q_H$;
\item the boundary condition $\L_D$ is compatible with~$Q_C$;
\item the boundary condition $\L_{(2,2),L}$ is compatible with both the $C$ and $H$ deformations for any Lagrangian $L \subset F$.
\end{itemize}
Further, consider the holomorphically twisted 3d $\N=4$ vector multiplet, with space of boundary fields $E_\partial = T^* \left( \Omega^{0,*}(\C)[1] \otimes \lie{g}[\epsilon] \right) $. Then:
\begin{itemize}[itemsep = 0 pt]
\item the boundary condition $\L_N$ is compatible with $Q_C$;
\item the boundary condition $\L_D$ is compatible  with $Q_H$;
\item the boundary conditions $\L_{(2,2),N}$ and~$\L_{(2,2),D}$ are compatible with both the $C$ and $H$ deformations.
\end{itemize}
\end{fact}
We note that our naming conventions for boundary conditions for the  vector multiplet are opposite to those in~\cite{CostelloGaiotto}; our convention refers to the boundary conditions on the \emph{scalars} in the vector multiplet, whereas that of~\cite{CostelloGaiotto} refers to the gauge field itself (which must have boundary conditions of the opposite type).

Let us give a more detailed example of such an inspection. We consider the case of the $Q_C = \epsilon \,\partial/\partial z$-twisted hypermultiplet. Here, we recall that $\L_N = \Pi F \otimes \Omega^{0,*}[1]$ and $\L_D = \eps \Pi F \otimes \Omega^{0,*}[1]$, with 
\deq{
\E_\bdy^C = \left( \L_N \xrightarrow{Q_C} \L_D \right).
}
Furthermore, we can identify $\L_N$ canonically with $\L_D^\vee$. We note that the inclusion map 
\begin{equation}
i: \L_D \rightarrow \E_\bdy^C, \qquad
\begin{tikzcd}
0 \ar[r] \ar[d] & \L_N \ar[d, "Q_C"] \\
\L_D \ar[r, "\text{id}"] & \L_D
\end{tikzcd}
\end{equation}
is a map of chain complexes, because the square commutes. On the other hand, the collection of maps
\[
\begin{tikzcd}
\L_N \ar[r, "\text{id}"] \ar[d] & \L_N \ar[d, "Q_C"] \\
0 \ar[r, "\text{id}"] & \L_D
\end{tikzcd}
\]
is \emph{not} a commuting square and does not define a subcomplex of~$\E_\bdy^C$.
As such, it is clear that $\L_D$ is a subcomplex of the $Q_C$-deformed space of boundary fields, whereas $\L_N$ is not.


\section{The brane perspective on deformable boundary conditions}
\label{sec:Three}

\par Having described the compatibility of the boundary conditions with either types of topological twists from the point of view of field theory using the BV formalism, we now turn to a string theory approach, for a complementary analysis. In this context, the bulk 3d theories are engineered as theories in the worldvolume of D3 branes suspended along a dimension (in the configurations at hand that would be $x^{6}$) between IIB fivebranes, as shown in Table~\ref{tab: branes}. In the IR, the finite distance becomes irrelevant and the the worldvolume theory is effectively three dimensional. Accordingly, the various boundary conditions are introduced by additional IIB five-branes, appropriately inserted in the brane configuration realising the bulk theory. In the following, starting from a configuration in the flat brane limit, we introduce the 3d theories with boundaries and comment on their field content. 
We examine whether the desired deformation of the preserved boundary supersymmetry can be captured by a geometric deformation of the branes introducing boundary conditions. 

\subsection{Three-dimensional $\N=4$ theories from flat-space string backgrounds}
\label{ssec:HW}

Three-dimensional $\N=4$ theories can be obtained in Type IIB string theory as the low-energy theory of configurations involving D3, D5 and NS5 branes in flat space \cite{HananyWitten} :
\begin{table}[h]
\begin{equation*}
\begin{array}{c || cccccccccc}
&0&1 & 2& 3& 4& 5& 6& 7& 8& 9\\
\hline
\text{D3}&\bullet&\bullet& \bullet& & & & \bullet& & & \\
\hline
\text{D5}&\bullet&\bullet&\bullet& & & & & \bullet&\bullet& \bullet \\
\hline
\text{NS5}&\bullet&\bullet & \bullet& \bullet& \bullet& \bullet& & & &
\end{array}
\end{equation*}
\caption{IIB brane configuration, engineering the 3d $\N=4$ theory. The strings suspended between the branes of the configuration provide gauge and flavour groups, as described in the text.}
\label{tab: branes}
\end{table}
With the insertion of the above branes, the initial $\SO(1,9)$ Lorentz symmetry breaks down to $\SO(1,2)_{[012]}\times \SO(3)_{[345]}\times \SO(3)_{[789]}$. We identify $\lie{so}(3)_{[345]}\times \lie{so}(3)_{[789]}$ with (the Lie algebra of) the $R$-symmetry group $\lie{sp}(C)\times \lie{sp}(H) = \lie{so}(C \otimes H)$; it is clear that $\SO(1,2)_{[012]}$ should be identified with the Lorentz group $\Spin(V) \cong \SU(S)$. It is straightforward to see that the configuration preserves eight real Poincar{\'e} supercharges, corresponding to $\mathcal{N}=4$ supersymmetry in three dimensions, transforming in the spinor representation of the Lorentz and R-symmetry groups. This configurations provide a particularly practical realization of three dimensional mirror symmetry as S-duality in type IIB string theory, under which the two types of fivebranes are exchanged while the D3 branes are retained. 

In the present work we will be interested in theories of vector multiplets or hypermultiplets and in their boundary conditions. Recall that the dimensional reduction of the D3-brane worldvolume theory to three dimensions is maximally supersymmetric Yang-Mills theory; in three dimensional $\N=4$ language, the fields consist of one vector multiplet and one hypermultiplet. The low energy worldvolume theory of D3 branes  suspended in the $x^6$ direction between either D5 or NS5 branes, is effectively three-dimensional, but contains only one of the two 3d $\N=4$ multiplets due to the additional boundary conditions; it is described by free 3d $\N=4$ hypermultiplets or vector multiplets respectively. 
In the first case, the Dirichlet boundary conditions imposed by the two D5 branes impose vanishing of the vector multiplet scalars. The hypermultiplet scalars correspond to the remaining worldvolume moduli, consisting of deformations in the $x^{7,8,9}$-directions (along the D5-branes), together with the vacuum expectation value of the scalar component $A_6$ of the gauge field on the D3 brane.

For a D3 suspended between two NS5 branes, the Neumann boundary conditions retain the vector multiplet scalars (transverse deformations in the $x^{3,4,5}$-directions) as remaining worldvolume moduli, resulting to a free vector multiplet theory. Of course, as reviewed above, the gauge field can be electric-magnetically dualized and viewed as a fourth scalar, so there is no contradiction with the fact that the moduli space is expected to be hyperk\"ahler.
On the contrary, in the same logic, the worldvolume theory of a number of D3 branes extended between an NS5 and a D5 brane has no massless moduli.

For consistency we remark that it is straightforward to construct  sQCD theories and general unitary linear quivers in this fashion \cite{Gaiotto:2008ak}: such theories arise from D3 branes suspended between NS5s and intersected by D5 branes. In such brane configurations, strings streched between two subsequent stacks of D3 branes, provide hypermultiplets transforming the in bi-fundamental representation of the adjacent gauge groups, whereas strings stretched between D3 and D5 branes provide hypermultiplets transforming in the fundamental representation of the gauge group. The aforementioned cases are found in Figure~\ref{fig:f1} below.
\begin{figure}[h]
\centering
  \includegraphics[width=0.8\linewidth]{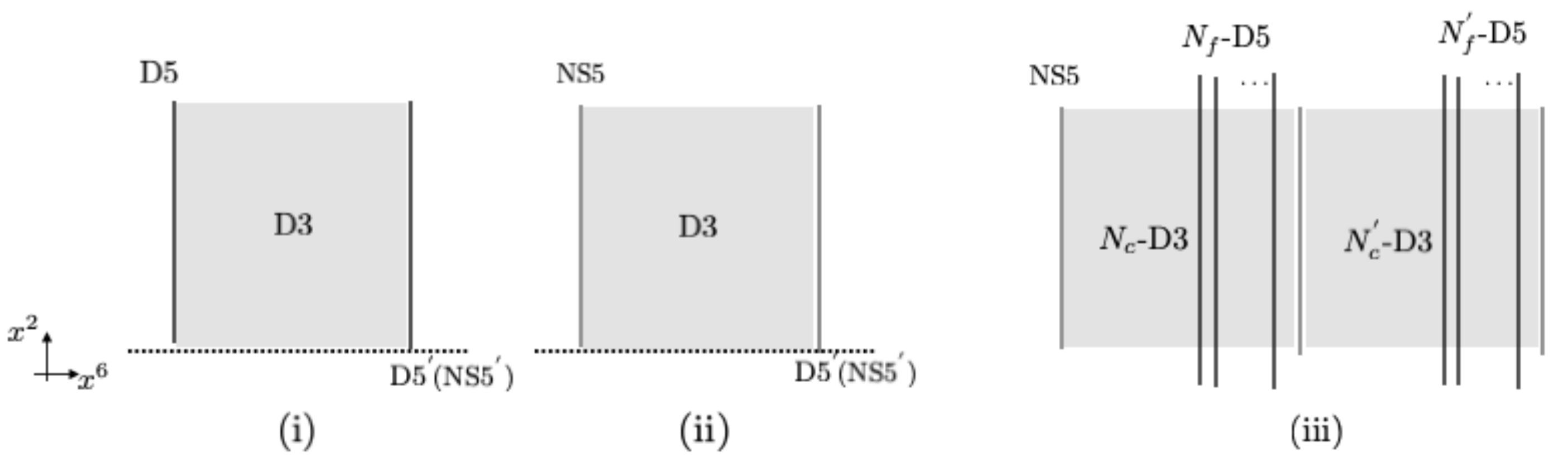}
  \caption{Brane configurations for a free hypermultiplet (i), for a free vector multiplet (ii) and for a quiver theory with gauge group $U(N_{c})\times U(N_{c}^{'})$ and flavour group $U(N_{f})\times U(N_{f}^{'})$}
  \label{fig:f1}
\end{figure}

\paragraph{Moduli space and brane moves} The moduli space of vacua of generic three dimensional $\N=4$ theories has the structure of a union of submanifolds parametrized by the vacuum expectation values (vevs) of the vector and hypermultiplet scalars, characterized as mixed-branches.  In the present work we focus on the two special subsets of these spaces. The Higgs branch, which is parametrized only by the vevs of the hypermultiplet scalars forming meson operators and the Coulomb branch, which is parametrized only by the vevs of vector mulltiplet scalars and in particular by the real scalar of the multiplet and abelian monopole operators.  These are non-compact hyper-K{\"a}hler manifolds with an $SU(2)_{H,C}$-action respectively and intersect at the origin, at the superconformal fixed point. A rigorous mathematical definition of the two branches has been obtained in \cite{Braverman:2016wma,Nakajima:2015txa,Nakajima:2017bdt}. Given a generic 3d theory $\N$=4 with gauge group $\mathbb{G}$ and $M$ a symplectic representation of $\mathbb{G}$, the Higgs branch can be formally defined as a hyper-K{\"a}hler quotient of $M$ by $\mathbb{G}_{c}\subset\mathbb{G}$: $M /\!/\!/ \mathbb{G}_{c}$, with $\mathbb{G}_{c}$ being a maximal compact subgroup of the gauge group. The Coulomb branch can be defined as an affine algebraic variety equipped with a holomorphic symplectic structure on the regular locus, together with a $\mathbb{C}^{*}$ action. At a generic point on the Higgs branch, the gauge group is completely broken and the IR theory is a free-hypermultiplet theory, whereas on the Coulomb branch the gauge group is broken to its maximal torus: $\mathbb{G}\hookrightarrow U(1)^{\rank(\mathbb{G})}$ and the IR theory is a theory of free vector multiplets. 
\par The dimension of either of the above branches of the moduli space can be easily read off the brane configuration realizing the theory as in \cite{Carta:2016fjb}. Coulomb and Higgs branch moduli correspond to motions of D3 brane segments moving along the NS5 and D5 directions respectively. A typical example which will also be used later is the one of the T[SU(2)] theory: it belongs to the category of T[SU(N)] theories, which are mirror-self dual, with $\text{SU(N)}_{f}$ flavor symmetry and $\text{U(1)}_{T}^{N-1}$ topological symmetry enhanced to SU(N) as an effect of the monopole operators. The theory is realized by a D3 brane suspended between two NS5 branes corresponding to a U(1) gauge group and intersected by two D5 branes realizing two fundamental hypermutiplets, as seen in figure below. By fragmenting the D3 brane into two segments stretching between each NS5 and the adjacent D5 and into one stretching between the two D5s. The former D3 branes have no massless moduli while the motion of the latter along the D5 directions results to Higgs branch moduli. By performing S-duality, or equivalently by moving the D5 branes on the right of the rightmost NS5 -taking into account the new D3 created by the Hanany--Witten transition- the the Coulomb branch moduli result from the D3 segment moving along the NS5 directions. Both branches are one-dimensional and the equality of dimensions is expected due to the self-duality that characterizes this class of theories. 
\begin{figure}[h]
\centering
  \includegraphics[width=0.45\linewidth]{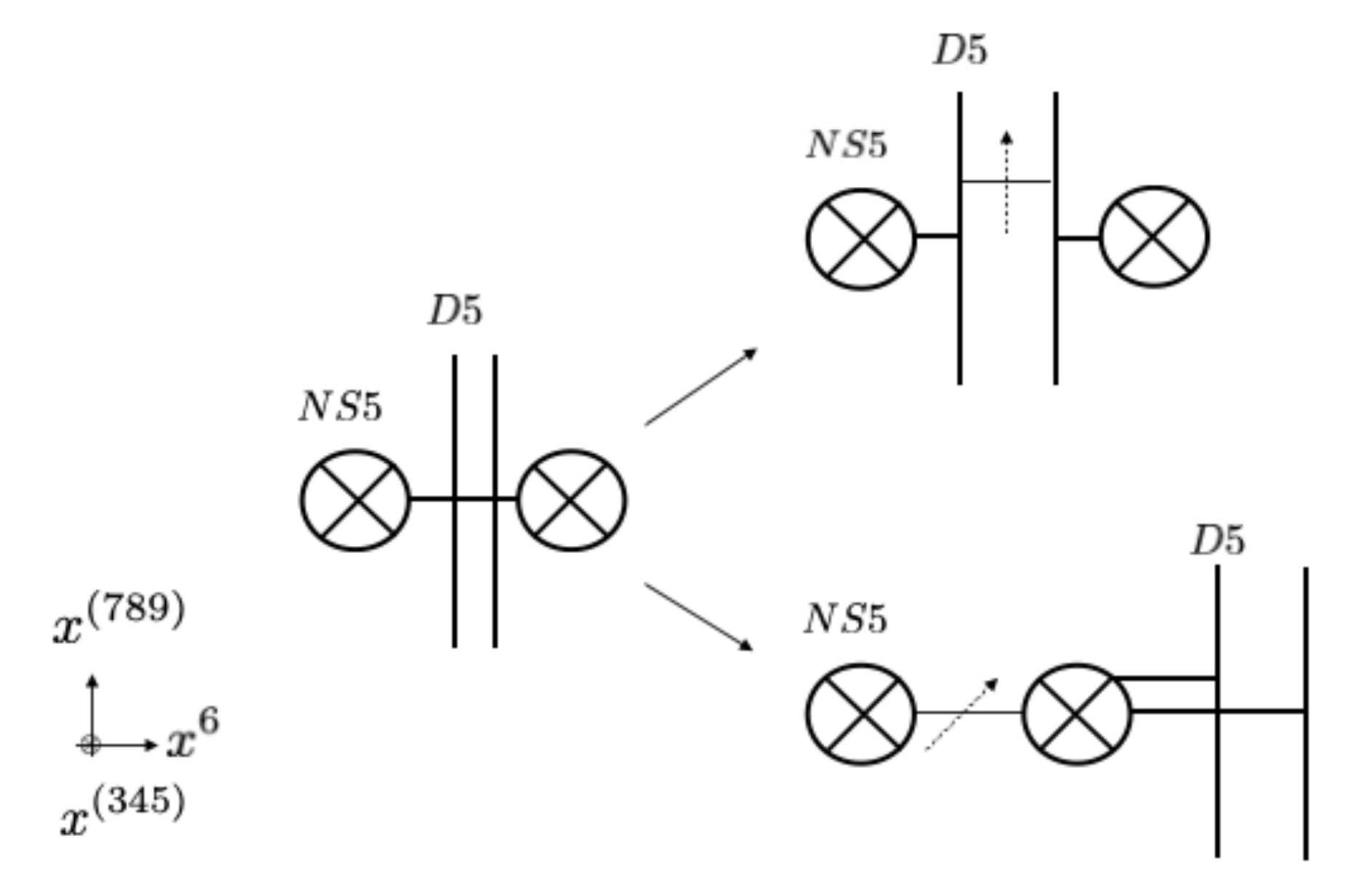}
  \caption{Higgs and Coulomb branch of the T[SU(2)] theory generated by D3 motions along the fivebrane directions. The brane diagram here is arranged so that the motions of D3 segments are visible. }
  \label{fig:f1}
\end{figure} 
We will return to the above example in section 5, in the context of twisted supergravity. 

 \subsection{Boundary conditions  from branes}

The two boundary conditions of 3d $\mathcal{N}=4$ theories relevant to the present study, are those preserving $\mathcal{N}=(2,2)$ and $\mathcal{N}=(0,4)$.
Both types have been extensively studied in \cite{Chung:2016pgt}, whereas the field theory and brane engineering of  2d $(0,4)$ boundary theories have been particularly studied in \cite{Hanany:2018hlz}. We review the construction of the boundary conditions and the fields surviving on the boundary in each case.

\paragraph{\N=(0,4) boundary conditions} In the standard 3d $\N=4$ brane configuration, one can appropriately introduce an additional five-brane fixing a boundary and preserving  $\N=(0,4)$ supersymmetry. The possible choices are either a D5 brane (designated D$5'$) extended along $(x^{0}, x^{1}, x^{3}, x^{4}, x^{5}, x^{6})$, or an NS5 (designated NS$5'$) brane occupying  $(x^{0}, x^{1}, x^{6}, x^{7}, x^{8}, x^{9})$. The chirality condition on the IIB Weyl spinors then reads $\Gamma_{013456}\tilde{\varepsilon}=\varepsilon$ and $\Gamma_{016789}\varepsilon=\varepsilon$, $\Gamma_{016789}\tilde{\varepsilon}=-\tilde{\varepsilon}$ respectively. Combined with the conditions imposed by the branes of the initial configuration, reduces the number of preserved supercharges to four and indicates the preserved amount of supersymmetry. 
\par Regarding the case of the vector multiplet, an extra D$5'$ brane imposes Dirichlet boundary conditions on the D3-worldvolume fields. The ones surviving on the boundary comprise a $(0,4)$ twisted hypermultiplet, with the complex scalars transforming singlets of $SU(2)_{H}$ and as spinors of $SU(2)_{C}$ and half of the initial fermions (right-moving, spinors under $SU(2)_{H}$). On the other hand the presence of an NS$5'$ brane imposes Neumann boundary conditions, with a $(0,4)$ vector multiplet on the boundary, that is, 2d gauge field and left moving fermions. 
\par The case of the hypermultiplet is similar. Dirichlet boundary conditions imposed by a D$5'$ brane,  complex fermions of the hypermultiplet survive on the boundary, comprising a $(0,4)$ Fermi multiplet. Finally, Neumann boundary conditions imposed by the insertion of an NS$5'$ leaves the complex scalars being $SU(2)_{C}$ singlets and transforming in the spinor of $SU(2)_{H}$ and half of the fermions on the boundary, forming a $(0,4)$ hypermultiplet.
\begin{table}[h]
\begin{equation*}
\begin{array}{c || cccccccccc}
&0&1 & 2& 3& 4& 5& 6& 7& 8& 9\\
\hline
\text{D}5'&\bullet&\bullet& & \bullet & \bullet & \bullet & \bullet & & & \\
\hline
\text{NS}5'&\bullet&\bullet&& & & &  \bullet& \bullet&\bullet& \bullet \\
\hline
\hline
\text{D}5''&\bullet&\bullet& & & &  \bullet& \bullet& \bullet & \bullet & \\
\hline
\text{NS}5''&\bullet&\bullet& & \bullet & \bullet &  & \bullet & & & \bullet \\
\end{array}
\end{equation*}
\caption{Additional fivebranes introducing boundaries for the 3d $\N=4$ theory. Introducing the D$5'$ or NS$5'$ branes results to a boundary reserving $\N=(0,4)$ supersymmetry, while in the case with D$5''$ or NS$5''$ the preserved boundary supersymmetry is $\N=(2,2)$.}
\end{table}
\hfill\break
\paragraph{\N=(2,2) boundary conditions} For reasons not only of consistency, but also for comparison with the case of interest, we present the brane engineering of $\N=(2,2)$ boundary condition.  As before, the corresponding brane configuration includes extra D5 or NS5 branes occupying  $(x^{0}, x^{1}, x^{5}, x^{6}, x^{7}, x^{8})$ and $(x^{0}, x^{1}, x^{3}, x^{4}, x^{6}, x^{9})$, respectively. 
\par Imposing Neumann boundary conditions to a bulk vector multiplet, reduces the boundary field content to a $(2,2)$ vector multiplet, consisting of a 2d vector and complex scalar. On the other hand, the two real scalar fields  surviving a Dirichlet boundary condition, for, a $(2,2)$ twisted chiral multiplet. Regarding the free hypermultiplet, the Neumann and Dirichlet boundary conditions pick the real and the imaginary part respectively,  of the two bulk complex scalars - hence sets of two real scalars -  each forming a $(2,2)$ chiral multiplet preserved on the boundary.

 \subsection{Topological twists from target-space geometry}

As we have already anticipated, the operation of applying a twisting homomorphism corresponds, in the context of theories on brane worldvolumes, to an appropriate choice of target space geometry. Considering D3 branes on a curved background, the effective theory in the D3 worldvolume can be topologically twisted by first defining the theory on three-manifold $M_{3}$ and consequently by embedding $M_{3}$ along an appropriate cycle in a special holonomy manifold. For example, the twisting homomorphisms $\phi_{C,H}$ discussed above for 3d $\N=4$ theories place the scalars that represent transverse deformations of the brane into one-form representations of~$\Spin(3)$. This can be accomplished by placing $M_{3}$ inside its cotangent bundle $T_{H,C}^{*}M_{3}$ along the zero section. In that case $M_{3}$ is embedded as a special Lagrangian submanifold inside a (noncompact) Calabi--Yau threefold, and the Killing spinor preserved by such a configuration corresponds to the scalar supercharge. We further note that the Weinstein Lagrangian neighborhood theorem~\cite{Weinstein} ensures that a tubular neighborhood of any Lagrangian cycle $L$ in a symplectic manifold is symplectomorphic to~$T^*L$, so that the global properties of the bulk geometry are not essential to the story.

Let us look a bit more explicitly into the example of the $C$-twist and the corresponding scalar supercharge, beginning from the brane configuration introduced in the previous section. Along with the chirality conditions $\Gamma\varepsilon=\varepsilon$ and $\Gamma\tilde{\varepsilon}=\tilde{\varepsilon}$ for the two 
Weyl spinors parametrizing the IIB string theory supersymmetries, the conditions imposed from the D3, D5 and NS5 branes of the configuration are given below:
\begin{align*}
&\text{D3}: \Gamma_{0126}\tilde{\varepsilon}=\varepsilon, &\text{NS5}: \Gamma_{012345}\tilde{\varepsilon}=\varepsilon \\
&\text{D5}:\Gamma_{012789}\tilde{\varepsilon}=\varepsilon, &\text{NS5}: \Gamma_{012345}\tilde{\varepsilon}=\varepsilon.
\end{align*}
Here $\Gamma$ and $\Gamma_{\mu}$ denote the chirality matrix and the ten-dimensional gamma matrices respectively. The twisting homomorphism $\phi_C$ is implemented as a twist of the Lorentz group by the $\Spin(3)_{C}$ subgroup of $SO(4)_{R}$, which is the diagonal embedding
 \deq{
 \phi_{C}: \Spin(3)_{L}\rightarrow \Spin(3)_{L}\times \Spin(3)_{C}.
 }
 Recalling that the Lie algebra of $\Spin(10)$ consists of antisymmetric quadratic expressions in the gamma matrices, and that there is a natural block-diagonal embedding
 \deq{
 \Spin(3)_L \times \Spin(3)_C \times \Spin(3)_H \hookrightarrow \Spin(10),
 }
 we observe that applying the twisting homomorphism just amounts to making the replacement
\begin{equation}
\Gamma_{\mu\nu}\rightarrow \Gamma_{\mu\nu}+\Gamma_{(\mu+3)(\nu+3)},\quad \mu,\nu=0,1,2
\end{equation}
in the above constraint equations.
It is further immediate that a supercharge is $\phi_C(\Spin(3))$-invariant (i.e., twists to a scalar) precisely when the conditions
\begin{equation}
(\Gamma_{\mu\nu}+\Gamma_{(\mu+3)(\nu+3)})\varepsilon=0,\quad \mu=0,1,2
\end{equation}
are satisfied.
The scalar supercharge is then identified using the projections imposed by the branes, the IIB chirality condition and the condition:
\begin{equation}
\Gamma_{\mu(\mu+3)}\varepsilon=\varepsilon.\quad \mu=0,1,2
\end{equation}
The resulting five independent equations constrain the 32 supersymmetries of the IIB theory to a one-component object which corresponds to the preserved scalar nilpotent supercharge picked up by the above twisting homomorphism. The second admissible fully topological supercharge is picked up by the twisting homomorphism that carries the Lorentz group to~$SU(2)_{H}$. 

We remark now that there is an important subtlety, due to what we have already pointed out above: the action of the $R$-symmetry on the 
fields of the hypermultiplet is not unique, but can mix nontrivially with the symplectic flavor symmetry. Recall the twisted field content listed above in Table~\ref{tab: twisted}. It is clear that the scalars in the $\phi_C$ twist of the vector multiplet are a one-form, whereas the scalars in the $\phi_H$ twist of the vector multiplet remain scalars in a three-dimensional representation of $R$-symmetry. The situation for the hypermultiplet, with the action of both $R$-symmetry and $\lie{sp}(F)$ flavor symmetry as defined above, is more subtle. $\phi_C$ again has no effect, and we are left with four scalars transforming in $H \otimes F$. But the $\phi_H$ twist leaves the scalars in the tensor product of the spin bundle with a rank-two symplectic bundle defined by~$F$.

So it is natural to ask if there is a class of manifolds of special holonomy, for which the deformations of a particular kind of supersymmetric cycle can be identified with the sections of such a bundle over the cycle itself. There is a natural candidate, given by manifolds of $G_2$ holonomy. Recall that, for a Calabi--Yau geometry, Weinstein's theorem gave an identification $NL \cong T^*L$ between the normal and cotangent bundles. Work of McLean~\cite{McLean} shows that, for $C$ a coassociative four-cycle in a $G_2$ manifold,
\deq{
NC \cong \wedge^2_-T^*C,
}
so that infinitesimal deformations are  identified with anti-self-dual two-forms. In each case, the deformations as a calibrated cycle are identified with the spaces of \emph{harmonic} forms of appropriate type, and are unobstructed; thus, things depend only on the topology of the cycle, and  not on the ambient manifold of special holonomy.

Since there are four scalar fields, though, we are interested in a three-dimensional calibrated cycle in a $G_2$ manifold, in other words an associative cycle. Indeed, it was proved in~\cite{McLean} that  the complexified normal bundle is isomorphic to the tensor product of the spin bundle with a particular rank-two symplectic vector bundle, and that deformations of ``harmonic twisted spinors,''  i.e.\ to the  kernel of the Dirac operator acting on sections of this bundle. (We note that this case is quite distinct from the others; for one thing, the space of deformations is not intrinsically determined just by the topology of the cycle itself. Moreover, infinitesimal deformations may fail to be unobstructed, and no cohomological argument can show that obstructions do not exist.)

It is moreover proved in~\cite{BryantSalamon} that a family of $G_2$ metrics exists on the total space of the (real rank four) spin bundle of any three-manifold with constant sectional curvature. (Since the spin representation of $\Spin(3)$ is complex, the complexification of the spin bundle consists just of the direct sum of two copies of the spin bundle, due to the well-known isomorphism
\[
\C \otimes_\R \C \cong \C \oplus \C;
\]
 the auxiliary bundle arising from the flavor space is thus trivial in this example.)
We can thus imagine two \emph{independent} geometric situations that give rise to twisting homomorphisms for the hypermultiplet, depending on whether or not the flavor symmetry is involved; one involves placing the theory on a Lagrangian cycle in a Calabi--Yau threefold (times \R), the other an associative cycle in a $G_2$ manifold. Clearly, the former is in a sense a special case of the latter.

\subsection{Geometric deformations of boundary branes}

Recall that we have considered boundary conditions of $\N=(0,4)$ type, and argued that, at least in certain cases, they define boundary conditions of  the holomorphic theory that are compatible with particular deformations to further topological twists. This implies that a family of deformations of such boundary conditions must exist, so that the family is compatible with the family of topological-type supercharges.
It is interesting to examine whether or not these deformations of the standard $\N=(0,4)$ boundary conditions can be obtained by geometrically deforming the boundaries by appropriate rotations of the boundary fivebranes. Through these geometric operations we are in particular looking for backgrounds preserving the $C$ or $H$ topological supercharges as deformations of a fixed holomorphic supercharge, labelled by some parameter of the geometry:

\begin{equation}
\mathcal{Q}_{C,H}^{\zeta}=Q_{\text{hol.}}+\zeta Q_{C,H}'.
\end{equation}
Here $Q_{\text{hol.}}$ is a (fixed) holomorphic supercharge that is left-moving  from the 2d $\N=(4,4)$ perspective, while $Q_{C,H}^{'}$ are right-moving supercharges. Let's consider first the $\N=(2,2)$ conditions, which preserve both the holomorphic and the deforming supercharges: the boundary conditions are automatically compatible with the holomorphic and topological twist:
\begin{equation}
Q_{\text{hol.}}\mathcal{B}=Q_{C,H}'\mathcal{B}=0.
\end{equation}

On the other hand, the pure $\N=(0,4)$ boundary conditions, while preserving the holomorphic supercharge, do not preserve the deforming supercharges, rendering them incompatible with the bulk topological twists:

\begin{equation}
Q_{\text{hol.}}\mathcal{B}=0,\quad ;\quad Q_{C,H}^{'}\mathcal{B}=\xi_{C,H}\neq 0
\end{equation}

Depending on the boundary condition, it may be the case that it can be made $Q_{\text{hol.}}$-exact, namely $\xi_{C,H}=Q_{\text{hol.}}\mathcal{O}_{C,H}$. Starting from the holomorphic boundary condition:
\begin{equation}
\mathcal{B}_{\text{hol.}}=(\mathcal{B},Q_{\text{hol.}}),
\end{equation}
it can be further deformed by the $C$ or $H$-type deforming supercharges in order to be compatible with the topological twist:
\begin{equation}
\mathcal{B}_{\zeta}=(\mathcal{B},\mathcal{O}_{\partial}+\zeta\mathcal{O}_{C,H}):\quad Q_{\zeta}^{C,H}\mathcal{B}_{\zeta}=(Q_{\text{hol.}}+\zeta Q^{C,H})\cdot(\mathcal{B},\mathcal{O}_{\partial}+\zeta\mathcal{O}_{C,H})=0
\end{equation}

Here we see how the deformation substantially cancels the failure of the boundary condition to be compatible with the $H$ and $C$-type deforming supercharges. The deformed boundary condition is now compatible with the topological supercharge $\mathcal{Q}_{C,H}^{\zeta}=Q_{\text{hol.}}+\zeta Q_{H,C}'$. 
As such, the family of boundary conditions we are looking for must have this property: the boundary condition $\mathcal{B}_\zeta$ must  preserve exactly $\mathcal{Q}_{C,H}^{\zeta}$, and not $Q_{\text{hol.}}$ and $Q_{C,H}'$ separately.

\paragraph{Boundary D5$'$ and NS5$'$ branes}
In the previous sections, we have thoroughly presented how an appropriate insertion of boundary five-branes introduces Dirichlet and Neumann boundary conditions for the bulk theory. The  2d boundary preserves $\N=(0,4)$ or $\N=(2,2)$ supersymmetry and it is the former case which can be deformed to be compatible with the bulk H or C-twist, while the latter is automatically compatible with both the holomorphic and the topological twist.
\par We proceed by using of the gamma matrix projection relations introduced in the previous subsection with the three- and five-branes occupying the indicated directions. Combined with the IIB chirality relations, lead to the following relations for the 4d subspaces:
\begin{equation}
\Gamma_{3456}(\varepsilon,\tilde{\varepsilon})=(\varepsilon,\tilde{\varepsilon})\quad ; \quad \Gamma_{6789}(\varepsilon,\tilde{\varepsilon})=(-\varepsilon,\tilde{\varepsilon})
\end{equation}

We can try to deform the $(0,4)$ Dirichlet boundary conditions introduced by the boundary $D5$ brane, by considering rotation of certain directions by an angle $\theta_{i}$. The projection reads:
\begin{equation}
\Gamma_{01}\mathcal{R}_{37}\mathcal{R}_{48}\Gamma_{56}\tilde{\varepsilon}=\varepsilon,
\end{equation}
where 
\begin{equation*}
\mathcal{R}_{\mu, \mu+4}=c_{i}\Gamma_{\mu}+s_{i}\Gamma_{\mu+4},\quad \mu=3,4,\quad i=1,2,
\end{equation*}
with $c_{i}=\text{cos}\theta_{i}$ and $s_{i}=\text{sin}\theta_{i}$. For $\theta_{1}=\mp \theta_{2}=\theta$ one obtains:
\begin{equation}
\varepsilon=c^{2}\Gamma_{01}\varepsilon \mp cs(\Gamma_{013568}\pm\Gamma_{014567})\tilde{\varepsilon}\mp s^{2}\Gamma_{015678}\tilde{\varepsilon}
\end{equation}

Imposing vanishing of the cross-term at finite $\theta$:

\begin{equation}
\Gamma_{013568}\tilde{\varepsilon}=\mp \Gamma_{014567}\tilde{\varepsilon}\to \Gamma_{3478}\tilde{\varepsilon}=\pm\tilde{\varepsilon},
\end{equation}

which results to the condition $\Gamma_{01}\varepsilon=\varepsilon$, leading back to $\N=(0,4)$ case. If instead we expand for small angle $\theta$ (take the case $\theta_{1}=\theta_{2}=\theta$) the expression results to:

\begin{equation}
\varepsilon=\Gamma_{01}\varepsilon+\theta(\Gamma_{48}-\Gamma_{37})\varepsilon+\theta^{2}(\mathbb{I}\varepsilon+\Gamma_{59}\tilde{\varepsilon})
\end{equation}

The right hand side commutes with $\Gamma_{01}$ and hence with the projectors $\mathbb{I}\pm\Gamma_{01}$, indicating that if a left mover is a solution, then also is separately the right-mover. In the above analysis, we where looking for $Q_{\zeta}$ as a solution for a value of $\zeta$ determined by geometric parameters; in the case at hand, determined by the angles between the various directions occupied by the boundary five branes. The computation indicates that this is in fact not possible: the holomorphic and the deforming supercharges separately solve the equation and the parameter $\zeta$ is undetermined. The studied configurations may either preserve the whole family ($\N=(2,2)$) or $\zeta=0$ ($\N=(0,4)$).

\paragraph{Boundary $(p,q)_{5}$ brane} Instead of a D$5'$ or an NS$5'$ brane one may consider the more general case of the $(p,q)_{5}$ brane occupying $(x^{0},x^{1},x^{3},x^{4},x^{5},x^{6})$. The projection relation, in the simplified case $\tau_{\text{IIB}}=i $, reads:

\begin{equation*}
\tilde{\varepsilon}=-\Gamma_{013456}(\lie{p}\tilde{\varepsilon}-\lie{q}\varepsilon),\quad \lie{p}=\frac{p}{\sqrt{p^{2}+q^{2}}}\quad\text{and}\quad\lie{q}=\frac{q}{\sqrt{p^{2}+q^{2}}}
\end{equation*}

Combining now with the D3 and D5 projections, results to:

\begin{equation*}
\tilde{\varepsilon}=\Gamma_{01}(\lie{q}\tilde{\varepsilon}-\lie{p}\varepsilon)
\end{equation*}

For $p=0, q=0$, this obviously boils down to the (01)-chirality relation for $\tilde{\varepsilon}$, as in the case of the boundary D$5'$ brane. An $x^{3}-x^{7}$ and $x^{4}-x^{8}$ rotation by opposite angles then gives:

\begin{equation*}
\Gamma_{01}\tilde{\varepsilon}=(c^{2}\lie{q}-s^{2}\lie{p}\Gamma_{59})\tilde{\varepsilon}-(c^{2}\lie{p}+s^{2}\lie{q}\Gamma_{59})\varepsilon-cs(\Gamma_{37}+\Gamma_{48})(\lie{q}\tilde{\varepsilon}-\lie{p}\varepsilon),
\end{equation*}

while expansion around small angles results to:

\begin{equation*}
\Gamma_{01}\tilde{\varepsilon}=(\lie{q}-\theta(\Gamma_{37}+\Gamma_{48})\tilde{\varepsilon}+(\lie{p}+\theta(\Gamma_{37}+\Gamma_{48})\varepsilon+\mathcal{O}(\theta^{2})
\end{equation*}

One can observe that also in this special case we do not obtain the deforming supercharge $Q_{\zeta(\theta)}$ as a solution, but rather either the holomorphic supercharge or a type-$H$ or $C$ deforming supercharge independently.

While having studied IIB brane configurations, we saw that we have a variety of different choices for the boundary branes. The choice of $(p,q)$-fivebrane is rather special. It is labelled by discrete parameters controlling the preserved supersymmetry (along with geometric parameters as angles in case of tilting). In M-theory, all distinctions of boundary fivebranes are geometrized and the choice is then restricted to a single object, which is the M5 brane. 
The claim in that case becomes more clear: we have already controlled a big family of geometric configurations of branes in IIB string theory, but none of these choices result in a $\zeta$-dependent family that preserves just the topological supercharge $\mathcal{Q}_{C,H}^{\zeta}=Q_{\text{hol.}}+\zeta Q_{H,C}'$ and thus captures the types of deformations introduced by Costello and Gaiotto. In the following section, we will see instead how we can bypass the question of deformability by working in twisted supergravity, where the holomorphically twisted theory and its boundary conditions can be directly geometrically engineered. This will allow us to recover criteria for compatibility of the $\N=(0,4)$ boundary conditions with topological twists from a new perspective.

\section{Deformable boundary conditions from twisted supergravity backgrounds}
\label{sec:IIB}

In the preceding sections, we have studied the notion of deformability of a boundary condition introduced in~\cite{CostelloGaiotto}. As recalled above, topological twists (of 3d $\N=4$ theories) can be viewed as further twists of the holomorphically twisted theory.\footnote{In general, any twist is a deformation of a minimal or holomorphic twist, where ``holomorphic'' must be understood appropriately in odd dimensions~\cite{NV}.} A boundary condition of a (three-dimensional $\N=4$) theory is \emph{deformable} when it is compatible with a holomorphic supercharge, and when its holomorphic twist is compatible with the action of the deforming supercharge on the holomorphic theory. Boundary conditions that are compatible with the topological supercharge in the untwisted theory (such as those of $\N=(2,2)$ type) are trivially deformable, but the converse implication is not true: examples are given by the boundary conditions of $\N=(0,4)$ type studied in holomorphic field theory above. The deformation that makes the boundary condition compatible with the topological supercharge is precisely the nullhomotopy that witnesses the exactness, with respect to the holomorphic supercharge, of the failure to be compatible with the deforming supercharge.

We went on to recall constructions due to Hanany and Witten that engineer the 3d $\N=4$ multiplets we are interested in using particular configurations of branes in type IIB supergravity theory (or string theory). Appropriate configurations of D3, D5, and NS5 branes can be used to arrive at three-dimensional theories of vector and hypermultiplets; introducing further D5 and NS5 branes engineers half-BPS boundary conditions of $\N=(0,4)$ or $\N=(2,2)$ type. Following~\cite{BershadskySadovVafa}, we reviewed how twisting homomorphisms can be incorporated in these constructions by an appropriate choice of (non-flat) target-space geometry. We also studied wide classes of geometric deformations of these setups, and argued that none of these deformations can induce the types of terms that we are interested in (those that create compatibility with a  bulk supercharge that is of topological type).

The discussion so far invites several natural questions. First and foremost, we emphasized in our discussion of twisting that two separate procedures are involved: a choice of twisting homomorphism, and the passage to cohomology of a square-zero supercharge (typically taken to be a scalar after applying the twisting homomorphism). In the previous section, following the work of~\cite{BershadskySadovVafa}, we recalled that twisting homomorphisms of brane worldvolume theories are related to different choices of the background geometries in which those branes sit. The second part of the twist construction---choosing to actually take invariants of the chosen supercharge---is not yet visible, and it is logical to ask if one might see this choice as related to a particular feature of the background geometry.

In our discussion of deformability, we have furthermore emphasized a two-step approach to the computation of a topological twist, which views it as a further deformation of a holomorphically twisted theory (at least on appropriately chosen spacetimes). This perspective was important in the insights into deriving holomorphic boundary conditions in topologically twisted theories from $\N=(0,4)$ boundary conditions in the work of~\cite{CostelloGaiotto}. 
Having understood, following~\cite{HananyWitten}, how such a 3d $\N=4$ theory can be constructed in type IIB string theory, it is natural to wonder if it is possible to understand the holomorphically twisted theory in analogous fashion, via some holomorphic form of geometric engineering. In fact, this question is not unrelated to the other one, since holomorphically twisted theories are essentially constructed by choosing to pass to invariants of a holomorphic supercharge; the twisting homomorphism plays a comparatively minor (though nonvanishing) role. 

Techniques for answering both questions are provided by the work of Costello and Li on twisted supergravity~\cite[for example]{CostelloLi}.
In this work, twisted supergravity backgrounds were defined to be supergravity backgrounds in which the (bosonic) ghosts of local supersymmetry acquire nonvanishing vacuum expectation values. It is then a matter of unpacking definitions to see that twisted supersymmetric field theories are nothing other than supersymmetric field theories coupled to a twisted supergravity background; this applies, in particular, to brane worldvolume theories. 

Having understood this, it remains only to come to an explicit understanding of the twisted theory of supergravity in relevant examples. For perturbative type IIB supergravity on flat space, an allowable vacuum configuration of the ghost field for local supersymmetry is just a constant square-zero supercharge. (In general, such nonzero configurations may or may not exist, depending on the metric, and are related to Killing spinors.) In particular, type IIB supergravity on flat space admits a holomorphic twist. Costello and Li made numerous conjectures about this holomorphic theory by leveraging information about the topological string; their essential insight was a conjectured identification between their version of BCOV~\cite{Bershadsky:1993cx} theory---the string field theory of the topological B-model---and holomorphically twisted type IIB string theory. Similarly, they identified a particular ``minimal'' subsector of this theory with holomorphically twisted type IIB supergravity. Further work has fleshed out these conjectures by providing more detail and showing how various aspects of string theory, such as $S$-duality, are realized at the level of the twist~\cite{SuryaYoo}.

Recently,  these conjectures have begun to be established by direct computations of twists of supergravity theories; see~\cite{spinortwist} for rigorous computations of the free holomorphic twists of the IIB, IIA, and eleven-dimensional supergravity theories that appeal to the pure spinor formalism, and~\cite{SuryaIngmarBrian} for a study of the interactions in holomorphic eleven-dimensional supergravity, as well as~\cite{EagerHahner} for an analysis of the maximal twist of free 11d supergravity from the component-field perspective, and the related work of Cederwall~\cite{Ced-towards,Ced-11d,Ced-SL5}.

For our purposes here, we will not need any detail about the interactions in BCOV theory, or about its quantization (though we emphasize that such results are available and could be applied; see~\cite{CLBCOV1,CLtypeI}.) We will offer some minimal review of the study of further deformations of holomorphic type IIB supergravity and branes in holomorphic supergravity backgrounds, following~\cite{CostelloLi}; the reader is referred to that paper for further information.

\subsection{Twisted supergravity and the topological $B$-model}

Costello and Li's version of BCOV theory is constructed as a holomorphic BV theory of the sort we have been discussing; it can be defined on any odd Calabi--Yau manifold, but we will only be interested in the case of fivefolds, as we wish to connect to supergravity theories in ten  dimensions. The fields of BCOV theory are defined using the sheaf of \emph{polyvector fields}, which is 
\deq{
\PV^{i,j}(X) = \Omega^{0,j}(X, \wedge^i TX).
}
Here $TX$ denotes the \emph{holomorphic} tangent bundle of the Calabi--Yau manifold; $\Omega^{0,0}(X, \wedge^i TX)$ is the space of smooth global sections of this bundle, for example. On flat space, the complex
\[
\PV^{i,\bu}(\C^d) = \left( \Omega^{0,\bu}(\C^d, \wedge^i T\C^d), \overline\partial\right)
\]
is the Dolbeault (smooth) resolution of the space of holomorphic sections of the bundle $\wedge^i T\C^d$. More generally, we think of $\left(\PV^{i,\bu}(X),\overline\partial\right)$ as a derived replacement for holomorphic $i$-polyvector fields  on~$X$ that may have higher cohomology.

For our purposes, a Calabi--Yau manifold is  equipped with  a holomorphic volume form $\Omega$ that witnesses the trivialization of the   canonical bundle. Contraction with the holomorphic volume form defines an isomorphism
\deq{
\Omega \vee \cdot : \PV^{i,j}(X) \xrightarrow{\cong} \Omega^{d-i,j}(X).
}
Pulling back the  holomorphic  de Rham  operator on $X$ under this isomorphism defines an operator $\del_\Omega$ that is a holomorphic analogue of the divergence operator. 

To write the space of fields of free BCOV theory on~$X$, we adjoin a parameter $t$ of cohomological degree 2, and consider the complex
\deq{
\E = \left( \PV^{\bu,\bu}(X)\lbr t \rbr [2], \overline\partial + t \partial_\Omega \right).
}
We observe that the shift $\E[-1]$ admits the structure of a local dg Lie algebra by equipping it with the Schouten bracket. While this dg Lie algebra structure is \emph{not} the local $L_\infty$ structure that corresponds to the BCOV action functional, it is closely related, and it will be sufficient for our purposes to consider just the dg Lie algebra structure arising from the Schouten bracket. 
While the space of fields is not $(-1)$-shifted symplectic, as it would be in a standard BV theory, the  space of observables  can be equipped with  a  mildly degenerate $(+1)$-shifted Poisson bracket, and much of the standard yoga of the Batalin--Vilkovisky formalism continues to apply. We refer  to~\cite{CostelloLi} for details.

At the level of the free theory, it is conjectured in~\cite{CostelloLi} and proved in~\cite{spinortwist} that the holomorphic twist of  type IIB supergravity on flat space is \emph{minimal} BCOV theory on~$\C^5$. (More precisely, \cite{spinortwist} proves that the holomorphic twist of type IIB supergravity is  a \emph{presymplectic} version of minimal  BCOV  theory, with the degeneracy coming from the self-dual four-form field of the type IIB multiplet; minimal  BCOV theory is a ``theory of field strengths'' for this multiplet.) Minimal BCOV theory on a $d$-fold consists of the subspace of fields lying in those summands $t^k \PV^{i,\bu}$ for which $i + k < d$. This subspace contains all of the propagating fields of the full theory; the remaining degrees of freedom consist purely of constant modes.

Since minimal BCOV theory is the holomorphic twist of type IIB supergravity, the holomorphic twist of the type IIB super-Poincar\'e theory should act on the theory on flat space. On general grounds, since local supersymmetry is gauged in supergravity theories, this action must be \emph{inner:} it should be witnessed by a map of Lie algebras from the global supersymmetry algebra to the local  $L_\infty$ algebra of the  fields of the interacting theory. This residual supersymmetry is computed in~\cite{CostelloLi}; in fact, it is the holomorphic twist of a particular central extension of the super-Poincar\'e  algebra that acts. Representation-theoretically, the odd elements of the twisted global algebra transform as one-forms and bivectors of the twisted Lorentz algebra~$\lie{sl}(5)$; their bracket is defined by contraction.
Under the map to the fields of minimal BCOV theory, these act by constant holomorphic bivectors in $\PV^{2,0}$ and linear holomorphic superpotentials in~$\PV^{0,0}$. It is trivial to check that the Schouten bracket pairs these to  even holomorphic vector fields in~$\PV^{1,0}$, which correspond to the inner action of the holomorphic translation symmetries of~$\C^5$.
(The central extension arises here because holomorphic translations have a nontrivial Schouten bracket with linear functions in~$\PV^{0,0}$; this central extension corresponds to the Noether charge of the fundamental string. We will not need to discuss it further in what follows.)

Costello and Li go on to construct a coupling of the fields of BCOV theory to holomorphic Chern--Simons theory. (In fact, they prove that BCOV theory is the \emph{universal} object that can be coupled appropriately to holomorphic Chern--Simons; \cite{CLBCOV1} shows that the  fields of BCOV theory are the cochain complex describing local first-order deformations of holomorphic Chern--Simons theory  by single-trace operators.) This coupling is the obvious map from polyvector fields on~$\C^5$ to endomorphisms of $\Omega^{0,*}(\C^5)$; after dimensional reduction, though, one must take some care to understand how the algebra acts on other holomorphically twisted D$(2k-1)$-branes. 
As we have repeatedly seen above, if a direction $z$ in a holomorphic theory is dimensionally reduced, the leftover odd generator $d\zbar$ of the Dolbeault complex survives as a  fermionic generator $\epsilon$. As such, we can  think of the theory on the holomorphic D$(2k-1)$-brane as (resolving) the sheaf of holomorphic functions on the supermanifold $\C^{k|5-k}$. The coupling map from BCOV theory to the reduced theory then takes the form
\deq{
z \mapsto \pdv{ }{\epsilon}, \qquad \pdv{ }{z} \mapsto \epsilon
} 
along the reduced directions, in keeping with expectations from the theory of Koszul duality. (See~\cite[Lemma~7.2.1]{CostelloLi}.)

\subsection{Holomorphically twisted brane constructions}

\subsubsection{The bulk complex structure and the D3 branes}
We build up the geometric engineering picture of boundary conditions in 3d $\N=4$ theories step by step in the setting of the holomorphic twist. For simplicity, we will just work with abelian theories, and as such do not consider interactions in any detail, other than as required to understand the inner action of twisted supersymmetry; we emphasize, however, that the interactions are understood and can be included in this setup. We follow the steps of the Hanany--Witten construction reviewed above, and show how these can be realized in a holomorphically twisted setting that explicitly geometrically engineers the field-theoretic results of~\S\ref{sec: FT}.

 We begin by just considering the holomorphically twisted theory on a single D3 brane. To couple it to BCOV theory, we must choose an appropriate complex structure. We will take the geometry to be of the form
 \begin{equation}
 \begin{tikzcd}[column sep = 0 pt, row sep = 2 pt]
 \text{bulk:}\quad & \C_{z_1} &\times& \C_{w_2}  &\times& \C_{z_2}  &\times& \C_{w_1}  &\times& \C_{z_3}  \\
  &\cong& &\cong& &\cong& &\cong& &\cong& &\\
 & \R^2_{01} &\times& \R^2_{26} &\times& \R^2_{34} &\times& \R^2_{59} &\times& \R^2_{78} \\ \hline
 \text{D3:}\quad & \C_{z_1} &\times& \C_{w_2} &\times& 0 &\times& 0 &\times& 0.
 \end{tikzcd}
\end{equation}
The theory in the bulk is BCOV theory on~$\C^5$; the theory on the worldvolume of the D3 brane is holomorphically twisted abelian 10d super Yang--Mills theory. As we have recalled above, the fields of this theory take the form
\deq{
\E = \Pi \Omega^{0,*}(\C^2)[\delta_1,\eps_2,\eps_3]
}
as a $\Z/2\Z$-graded BV theory. Here $\delta_1$, $\eps_2$, and~$\eps_3$ are odd parameters. In the B-model approach, this result can be recovered just by considering the open-string Hilbert space $\Ext^\bu_{\O_\C^5}(\O_{\C^2},\O_{\C^2})$. 
If we imagine constructing the theory by dimensional reduction, $\delta_1$ can be identified with~$d\bar{w}_1$ and $\eps_{2,3}$ with~$d\bar{z}_{2,3}$; this is the reason for our choice of notation, and the asymmetry will play a role in understanding the $\Z$-grading on the theory. To do this, we must appeal to a choice of twisting homomorphism in four dimensions, since the (interacting) ten-dimensional theory is not $\Z$-graded. Various choices are possible; we refer the reader to~\cite{ESW} for details. A choice which is convenient for our purposes, and matches with our conventions above for six-dimensional $\N=1$ theories, is to assign degree and parity $(1,+)$ to $\delta_1$ and $(0,-)$ to~$\eps_1$ and~$\eps_2$. The fields are then
\deq{
\E = \Omega^{0,*}(\C^2)[\delta_1,\eps_2,\eps_3][1]
}
The result is a $\Z\times \Z/2$-graded BV theory with a symplectic structure of bidegree $(-1,+)$. Another possible choice is to assign integer grading $+1$ to~$\epsilon_2$ and $-1$ to~$\epsilon_3$; the theory is then concentrated in even intrinsic parity, and one does not need to remember the $\Z/2$ grading explicitly. 

\subsubsection{Four-dimensional boundary conditions}
We now proceed to consider boundary conditions for D3 branes that arise by having them end on D5 or NS5 branes, as in the constructions in~\cite{HananyWitten}. (We do not yet consider a ``sandwich'' construction, but build up to this by first considering the relevant boundary conditions.) The relevant brane geometries are shown in Figure~\ref{fig:HWbdy}.
 \begin{figure}
 \begin{equation*}
 \begin{array}{c}
 \text{hypermultiplet:} \\
 \begin{tikzcd}[column sep = 0 pt, row sep = 2 pt]
 \text{bulk:}\quad & \C_{z_1} &\times& \C_{w_2}  &\times& \C_{z_2}  &\times& \C_{w_1}  &\times& \C_{z_3}  \\
  &\cong& &\cong& &\cong& &\cong& &\cong& &\\
 & \R^2_{01} &\times& \R^2_{26} &\times& \R^2_{34} &\times& \R^2_{59} &\times& \R^2_{78} \\ \hline
 \text{D3:}\quad & \C_{z_1} &\times& \H_{w_2} &\times& 0 &\times& 0 &\times& 0. \\ \hline
 \text{D5:}\quad & \C_{z_1} &\times& \R_2 &\times& 0 &\times& \R_9 &\times& \C_{z_3}
 \end{tikzcd}
 \end{array}
\end{equation*}
 \begin{equation*}
 \begin{array}{c}
 \text{vector multiplet:} \\
 \begin{tikzcd}[column sep = 0 pt, row sep = 2 pt]
 \text{bulk:}\quad & \C_{z_1} &\times& \C_{w_2}  &\times& \C_{z_2}  &\times& \C_{w_1}  &\times& \C_{z_3}  \\
  &\cong& &\cong& &\cong& &\cong& &\cong& &\\
 & \R^2_{01} &\times& \R^2_{26} &\times& \R^2_{34} &\times& \R^2_{59} &\times& \R^2_{78} \\ \hline
 \text{D3:}\quad & \C_{z_1} &\times& \H_{w_2} &\times& 0 &\times& 0 &\times& 0. \\ \hline
 \text{NS5:}\quad & \C_{z_1} &\times& \R_2 &\times& \C_{z_2} &\times& \R_5 &\times& 0
 \end{tikzcd}
 \end{array}
\end{equation*}
\caption{Boundary conditions arising from D3 branes ending on D5 or NS5 branes in twisted supergravity}
\label{fig:HWbdy}
\end{figure}

In order to introduce the branes that define boundary conditions, we are forced to deform our IIB background by a further twist. This is necessary because the support of the branes is not along a complex submanifold of~$\C^5$, as is already apparent because it defines a boundary condition for the D3 branes with real codimension one. Instead, the support of the branes is along the product of a complex submanifold of~$\C^3$ (in the $z$ directions) and a Lagrangian submanifold of~$\R^4$ (in the $w$ directions). The appropriate deformation is thus by the bivector $\del_{w_1} \wedge \del_{w_2}$. We end up in a twist where three translations survive. It is argued in~\cite{CostelloLi} that such a twist can be understood as a product of topological $A$ and $B$ models. The  deformation is well-defined on any product of a Calabi--Yau threefold and a Calabi--Yau twofold.

The deformation induces the half-topological twist of the worldvolume theory, in which the fields are 
\deq{
\E = \Omega^{0,*}(\C) \otimes \Omega^*(\R^2) [\eps_2,\eps_3][1].
}
This is now topological normal to the boundary; the boundary fields are the (unshifted) symplectic space
\deq{
\E_\bdy = \Omega^{0,*}(\C_{z_1}) \otimes \Omega^*(\R_2) [\eps_2,\eps_3][1].
}
The deforming element is $\delta_1 \partial/\partial_{w_2}$.

There are four obvious Lagrangians in this space: they correspond to the subspaces $\C[\eps_2]$, $\C[\eps_3]$, $\eps_3 \C[\eps_2]$, and $\eps_2 \C[\eps_3]$. The first two impose Neumann boundary conditions on the gauge field, and the last two Dirichlet boundary conditions on the gauge field. So they correspond to either NS-brane or D-brane boundary conditions respectively. The choice of one of the two is determined by the direction that the brane sticks out in; note the action of $\lie{sp}(1)$, which is the holomorphic twist of~$\lie{so}(5)=\lie{sp}(2)$. 
We will make a choice here, and consider the two boundary conditions defined by the Lagrangian submanifolds
\deq{
\L_{\text{NS5}} = \Omega^{0,*}(\C_{z_1}) \otimes \Omega^*(\R_2) [\eps_2][1] \subset \E_\bdy, \qquad 
\L_{\text{D5}} = \eps_3 \Omega^{0,*}(\C_{z_1}) \otimes \Omega^*(\R_2) [\eps_2][1]\subset \E_\bdy.
}
If we use the twisting homomorphism considered above that makes the theory integrally graded, then $\L_{\text{NS5}}$ contains copies of $\Omega^{0,*}(\C_{z_1}) \otimes \Omega^*(\R_2)$ starting in degrees $-1$ and $0$, whereas $\L_{\text{D5}}$ starts in degrees $-2$ and~$-1$.

\subsubsection{The Hanany--Witten construction: 3d $\N=4$ theories}

We now go on to consider  ``sandwiches'' of D3-branes stretched along an interval between two identical boundary conditions. On general grounds, the theory on the sandwich is the derived intersection of the Lagrangians defining the boundary conditions. Derived Lagrangian intersections in an unshifted symplectic space are $(-1)$-shifted symplectic~\cite{PTVV}. The derived self-intersection of a Lagrangian $\L$ is just $T^*[-1]\L$. 
To see this, it is easiest to recall that derived intersections are (at least locally) computed by taking the derived tensor product of functions on each Lagrangian over the ring of functions  on the  ambient  space:
\deq{
\O( \L \cup_X \L' ) =  \O(\L) \otimes^{\mathbb{L}}_{\O(X)} \O(\L').
}
To compute the  derived tensor product,  we need to freely resolve  $\L$ in modules over the ambient space. For the case we are interested in, this is straightforward: inside of $T^*\L \cong \L \oplus \L^\vee$, $\O(\L)$ has a straightforward free resolution by considering the Koszul complex $\O(\L \oplus \L^\vee \oplus \L^\vee[-1])$. Tensoring over~$\O(\L \oplus \L^\vee)$ with~$\O(\L)$, we obtain $\O(\L \oplus \L^\vee[-1])$ with zero differential, as claimed.
(This can also be seen by thinking about the BV formalism in terms of the derived intersection of the zero section with $\graph(dS)$ inside of the cotangent bundle to the field space; in that context, setting the action functional to zero returns the result of interest here.)

It is then obvious that we get the hypermultiplet for D5-branes and the vector multiplet for NS5-branes. If we use the integral grading considered above, then $T^*[-1]\L_\text{NS5}$ consists of four copies of holomorphic-topological functions on~$\C \times \R$, sitting in degrees $-1$ and $0$ in~$\L_\text{NS5}$, with duals in degrees $0$ and $-1$ respectively. This precisely matches the twist of the vector multiplet computed above.

On the other hand, $T^*[-1]\L_\text{D5}$ has four copies of holomorphic-topological functions, but in degrees $-2$ and~$-1$ (from $\L_\text{D5}$) with duals in degrees $+1$ and~$0$. This matches the twist of the hypermultiplet, after choosing a polarization and correspondingly making a choice of integral grading for which $\Pi F$ is replaced by $\C[1]\oplus\C[-1]$.
Note, though, that~$\L_\text{D5}$ is not the space of BRST fields of the hypermultiplet; the Lagrangian splitting of the BV theory that appears here is a different one, corresponding to the choice of a Lagrangian polarization in the flavor space $F$ (and matching the choice of $\epsilon_2$ versus $\epsilon_3$ appearing in defining $\L_\text{D5}$).

\subsection{Holomorphic boundary conditions in three dimensions}

Continuing our step-by-step discussion of holomorphic geometric engineering, we now introduce the additional branes, called D$5'$ or NS$5'$ above, that are responsible for boundary conditions in the 3d $\N=4$ theory. Recall that one expects half-BPS boundary conditions, both of $\N=(0,4)$ type and of~$\N=(2,2)$ type, to be compatible with the holomorphic twist. As such, we should not expect any further deformation of the IIB background to be required, and the branes (with support as introduced above) should define appropriate branes in the twisted theory. Indeed, it is simple to see that this is the case.

\subsubsection{Boundary conditions of $\N=(0,4)$ type}
Recall that Dirichlet boundary conditions for the 3d hypermultiplet were produced by D$5'$ branes sitting in the directions $013456$, giving rise to a boundary condition in the $2$ direction for the theory on the D3s. Similarly, Neumann boundary conditions were engineered by adding NS$5'$ branes along the $016789$ directions. Inserting these branes, we summarize the geometric setup in the following diagram:
 \begin{equation}
\qquad\qquad
 \begin{tikzcd}[column sep = 0 pt, row sep = 2 pt]
 \text{bulk:}\quad & \C_{z_1} &\times& \C_{w_2}  &\times& \C_{z_2}  &\times& \C_{w_1}  &\times& \C_{z_3}  \\
  &\cong& &\cong& &\cong& &\cong& &\cong& &\\
 & \R^2_{01} &\times& \R^2_{26} &\times& \R^2_{34} &\times& \R^2_{59} &\times& \R^2_{78} \\ \hline\hline
 \text{D3:}\quad & \C_{z_1} &\times& (\R_+)_2 \times I_6 &\times& 0 &\times& 0 &\times& 0. \\ \hline
 \text{\llap{hypermultiplet --- \quad}D5:}\quad & \C_{z_1} &\times& \R_2 \times \bdy I_6 &\times& 0 &\times& \R_9 &\times& \C_{z_3} \\ 
 \text{\llap{vector multiplet --- \quad}NS5:}\quad & \C_{z_1} &\times& \R_2 \times \bdy I_6 &\times& \C_{z_2} &\times& \R_5 &\times&0 \\ 
\hline\hline
 \text{\llap{$\N=(0,4)$ type --- \quad}
 D$5'$:}\quad & \C_{z_1} &\times& 0 \times \R_6 &\times& \C_{z_2} &\times& \R_5 &\times& 0 \\ \hline
 \text{
 NS$5'$:}\quad & \C_{z_1} &\times& 0 \times \R_6 &\times& 0 &\times& \R_9 &\times& \C_{z_3}
 \end{tikzcd}
 \label{eq:0,4support}
\end{equation}
Just as the original D3 and D5 branes do, both D$5'$ and NS$5'$ branes define products of complex submanifolds of~$\C^3_{z}$ and Lagrangian submanifolds of~$\R^4_w$. This is a legal brane setup in the IIB background we are considering, and no further deformation is required. 
As recalled above, the same new branes can be used to introduce boundary conditions of $\N=(0,4)$ type for the vector multiplet.
\subsubsection{Boundary conditions of $\N=(2,2)$ type}
Here, again, we introduce boundary conditions for the hypermultiplet and vector multiplet by appropriately including new branes of the types called D$5''$ and NS$5''$ above. Recall that the D$5''$ branes were supported along the directions $015678$ and the NS$5''$ branes along $013469$. 
It is easy to draw a diagram of the brane configuration in the holomorphic theory, which makes it clear that the branes inducing these boundary conditions are also compatible with the background and require no further deformation:
 \begin{equation}
 \qquad\qquad
 \begin{tikzcd}[column sep = 0 pt, row sep = 2 pt]
 \text{bulk:}\quad & \C_{z_1} &\times& \C_{w_2}  &\times& \C_{z_2}  &\times& \C_{w_1}  &\times& \C_{z_3}  \\
  &\cong& &\cong& &\cong& &\cong& &\cong& &\\
 & \R^2_{01} &\times& \R^2_{26} &\times& \R^2_{34} &\times& \R^2_{59} &\times& \R^2_{78} \\ \hline\hline
 \text{D3:}\quad & \C_{z_1} &\times& (\R_+)_2 \times I_6 &\times& 0 &\times& 0 &\times& 0. \\ \hline
 \text{\llap{hypermultiplet --- \quad}D5:}\quad & \C_{z_1} &\times& \R_2 \times \bdy I_6 &\times& 0 &\times& \R_9 &\times& \C_{z_3} \\ 
 \text{\llap{vector multiplet --- \quad}NS5:}\quad & \C_{z_1} &\times& \R_2 \times \bdy I_6 &\times& \C_{z_2} &\times& \R_5 &\times&0 \\ \hline\hline
 \text{\llap{$\N=(2,2)$ type --- \quad}
 D$5''$:}\quad & \C_{z_1} &\times& 0 \times \R_6 &\times& 0 &\times& \R_{5} &\times& \C_{z_3} \\ \hline
 \text{
 NS$5''$:}\quad & \C_{z_1} &\times& 0 \times \R_6 &\times& \C_{z_2} &\times& \R_9 &\times& 0
 \end{tikzcd}
 \label{eq:2,2support}
\end{equation}
Inspecting this diagram also makes it clear that the boundary conditions of $\N=(2,2)$  type  are a mixture of Dirichlet and Neumann boundary conditions for the scalars in  each  multiplet. 

\subsection{Further twists and compatibility}
\subsubsection{Residual supercharges as deformations of the supergravity background}
We recall that further twists of the IIB background we are working in are induced by either linear superpotentials or bivectors that are compatible with $\del_{w_1} \wedge \del_{w_2}$. According to~\cite{CostelloLi}, the map of residual supersymmetries of the background to deformations of the theory on the D3-branes is given by
\begin{equation}
\begin{gathered}
z_1, w_2 \mapsto z_1,w_2; \quad \pdv{ }{z_1}, \pdv{ }{w_2} \mapsto  \pdv{ }{z_1}, \pdv{ }{w_2} ; \quad \\
w_1,z_2, z_3 \mapsto \pdv{ }{\delta_1}, \pdv{ }{\epsilon_2}, \pdv{ }{\epsilon_3}; \quad
\pdv{ }{w_1}, \pdv{ }{z_2}, \pdv{ }{z_3} \mapsto \delta_1, \epsilon_2, \epsilon_3.
\end{gathered}
\end{equation}
More generally, coordinates and derivatives in directions along the worldvolume of a  brane map  in  the  obvious way; transverse coordinates map to fermionic derivatives, while transverse derivatives map to fermionic coordinate functions. It is important to emphasize that not all of these deformations can be interpreted as coming from the action of residual supersymmetries in the worldvolume theory. 
It remains for us to locate a copy of the twisted 3d $\N=4$ algebra within the space of further deformations; it will be important that the elements we identify act as supersymmetry transformations from the point of view of the branes (D3 and  D5 or NS5) that  engineer the 3d $\N=4$ theory. 

Recall that the twisted 3d $\N=4$ algebra contains two odd elements, $Q_C$ and~$Q_H$, which anticommute to the single surviving translation $\partial_{z_1}$. 
We further argued above that $Q_C$ acts in the twisted theory via a de Rham type differential, and $Q_H$ via a cancelling differential.

From the above, we can identify two pairs of candidate elements that act appropriately on the D3-brane worldvolume theory: we could choose the bivector $\partial_{z_1} \wedge \partial_{z_2}$  (which acts by the de Rham type differential $\epsilon_2 \partial_{z_1}$) and the linear superpotential $z_2$ (which acts by the fermionic derivative $\partial/\partial \epsilon_2$). Alternatively, we could choose $\partial_{z_1} \wedge \partial_{z_3}$ and~$z_3$, which also have the correct commutation relations. All of these are clearly compatible with the original deformation $\partial_{w_1} \wedge \partial_{w_2}$ that defined our background.

To distinguish between the two, we need to appeal to the way that the chosen elements act on the worldvolume theory of the D5 branes in our configuration. Since the $z_2$ direction is transverse to their support, it is clear that the elements $\partial_{z_1} \wedge \partial_{z_2}$ and~$z_2$ also act by residual supersymmetries there, by the same expressions as for the D3 branes. On the other hand, neither of the elements $\partial_{z_1} \wedge \partial_{z_3}$ or~$z_3$ act by a residual supersymmetry; the former induces a noncommutative deformation of the theory, while the latter deforms it to a curved theory in which the equations of motion (for a single brane) have no solutions at all. 
From these considerations, it follows immediately that, as deformations of the supergravity background, we should identify
\deq{
Q_C = \partial_{z_1} \wedge \partial_{z_2}, \qquad Q_H = z_2.} 
We note that it is more difficult to understand how the deformations act on the worldvolume theory of the NS5 brane, since an explicit coupling is not provided in~\cite{CostelloLi} in this case. However, it is reasonable to postulate in this nonminimally twisted background (just by consistency of the Hanany--Witten setup) that bivectors fully along the worldvolume of the NS5 brane and linear superpotentials associated to worldvolume coordinates act by supersymmetries in this case. There are two such linear superpotentials  $z_1$ and~$z_2$ and one such bivector $\partial_{z_1} \wedge \partial_{z_2}$; contrast this with the case of the D5 brane, where the supersymmetries would act by $\epsilon \partial_{z_1}$, $\epsilon \partial_{z_2}$, and $\partial/\partial \epsilon$. In either case, the commutation relations are the same, although the representation theory differs by a factor of the Calabi--Yau form.  

We can see that this identification is reasonable in a variety of ways. One is to recall that the fermionic coordinate appearing in the differentials that act in three dimensions must arise by dimensional reduction from six dimensions. In the Hanany--Witten picture, we can arrive at a consistent six-dimensional picture only by applying $T$-dualities along the NS5-brane worldvolume (and transverse to the D5-branes); as such, we identify the reduced directions with the $345$ directions, and the fermion with $\epsilon_2 = d\zbar_2$.

Another consistency check notes that the $C$ deformation makes $z_2$ topological and leaves $z_3$ holomorphic, whereas $Q_H$ makes $\C_{z_3}$ topological and leaves $z_2$ as a  holomorphic coordinate. Thinking about the identifications of $R$-symmetry makes it clear that this is sensible: we identified rotations in the $345$ directions with $\lie{sp}(C)$, and rotations in the $789$ directions with~$\lie{sp}(H)$. As such, $\lie{gl}(\C_2)$ can be identified with the Cartan subalgebra of~$\lie{sp}(C)$ and $\lie{gl}(\C_3)$ with the Cartan subalgebra of~$\lie{sp}(H)$.

We know that the $C$ twist should preserve those moduli that transform under the (untwisted) $H$ $R$-symmetry, and vice versa. 
In the geometric engineering picture, moduli are the positions of branes in either $\C_{z_2}$ or $\C_{z_3}$. When a transverse direction becomes topological in a given twist, the corresponding moduli are $Q$-exact (hence invisible) in the field theory description. For example, positions of branes in the $z_3$ direction become trivial in the $H$-twist, whereas they remain as interesting operators in the $C$-twist.
We will consider an example of this---the theory $T[SU(2)]$---in the following section, but we emphasize that the reasoning is perfectly general. 

\subsubsection{Compatibility and holomorphicity}

We now examine the compatibility of various boundary conditions with the deformations $Q_C$ and~$Q_H$, beginning with the D$5'$ and D$5''$ branes, where we  can explicitly understand the resulting deformations of the worldvolume theory.

Let us begin by examining the D$5''$ brane, which should define a boundary condition of $\N=(2,2)$ type that is compatible with both the $C$ and $H$ deformations, and which supports a topological theory in each case. In twisted supergravity, this should amount to seeing that the relevant deformations act by residual supersymmetries from the perspective of the D$5''$ worldvolume theory. Indeed, this is the case: we have 
\[
Q_C =  \pdv{ }{z_1} \wedge \pdv{ }{z_2} \mapsto \epsilon_2 \pdv{ }{z_1}, \quad
Q_H = z_2 \mapsto \pdv{ }{\epsilon_2},
\]
just as in the worldvolume theory of the D5 branes.

We go on to consider the D$5'$ brane, which defines boundary conditions of $\N=(0,4)$ type: Dirichlet  for the scalars in the hypermultiplet, or Neumann for the scalars in the vector multiplet, as can easily be read off from equation~\eqref{eq:0,4support}. Here, we would not expect either deformation to act by a  worldvolume supersymmetry, and indeed this is the case: following~\cite{CostelloLi}, $Q_C$ becomes the holomorphic Poisson bivector $\partial_{z_1} \wedge \partial_{z_2}$, and defines a noncommutative deformation of the worldvolume theory, whereas $Q_H$ becomes a ``curved'' deformation of the action by terms linear in the worldvolume fields. In this theory, the equations of motion for a single brane have no solutions at all. 

As such, it is pleasant to note that Neumann boundary conditions for the vector, or Dirichlet boundary conditions for the hyper, are obviously incompatible with the $Q_H$ deformation, in the sense that the equations of motion for the corresponding brane configuration have no solutions at all. On the other hand, the Poisson deformation does not eliminate any degrees of freedom whatsoever; moreover, the intersection of the D3 and D$5'$ branes (along $z_2 = x_5 =  0$) is isotropic with respect to the holomorphic symplectic form, so that one expects to see degrees of freedom supported on the boundary that are isomorphic to holomorphic functions of~$z_1$. This corresponds nicely to the explicit form of the boundary conditions as seen in field theory above.

We remark briefly on interpretations of these twisted supergravity backgrounds in terms of topological string theory. Costello and Li conjecture that the background of type IIB string theory where both of the deformations $\partial_{w_1} \wedge \partial_{w_2}$ and $\partial_{z_1} \wedge \partial_{z_2}$ are turned on is isomorphic (at least at the perturbative level) to a product of the topological $A$-model on~$\R^8 = \C^4_{z_1,z_2,w_1,w_2}$ and the topological $B$-model on~$\C_{z_3}$. We should thus expect that all of the branes that appear in our configuration define products of $A$-branes in~$\C^4$ and $B$-branes in~$\C_{z_3}$, The latter condition is obvious, since all branes are supported either along all of~$\C_{z_3}$ or at points. Referring to the tables in~\eqref{eq:0,4support} and~\eqref{eq:2,2support}, we see that  the D3 branes, as well as the D5 and D$5''$ branes, have half-dimensional support in~$\R^8$ and can thus be identified with Lagrangian objects of the  $A$-brane category. 

The situation is somewhat different for the D$5'$ branes, whose support is six-dimensional in~$\R^8$ and thus cannot possibly be Lagrangian. Since we are dealing with the $A$-model on a Calabi--Yau fourfold, though, there can be coisotropic objects in the $A$-brane category~\cite{KapustinOrlov}. Such a brane is defined by a foliation with a transverse holomorphic structure; at least in the hyperk\"ahler case, one expects there to be examples of such objects whose algebras of endomorphisms in the $A$-brane category are the deformation quantization of the algebra of holomorphic functions (with respect to the transverse holomorphic structure and a corresponding holomorphic symplectic form). 

It is pleasing that the space of fields we see matches what one would expect for such an object. Prior  to the deformation by the holomorphic Poisson bivector, the space of BV fields is
\[
\Omega^{0,\bu}(\C^2_{z_1,z_2}) \otimes \Omega^\bu_\text{dR}(\R^2_{56})[\epsilon_3][1],
\]
where a consistent $\Z$-grading is obtained by placing $\epsilon_3$ in  degree $-1$.  
The noncommutative deformation just deforms the commutative algebra structure by the holomorphic Poisson bivector.

\subsection{More complicated examples}

Given the techniques we have developed, it is straightforward to start studying more complicated examples. We will look at just one further example here, namely the $T[SU(2)]$ theory whose brane realization was discussed above in~\S\ref{ssec:HW} (see Figure~\ref{fig:f1}), 
The realization involves D3 branes stretched between a sequence of NS5 and D5 branes. We label the positions of the branes in the $\R_6$ direction by $a$, $b$, $c$, and~$d$; the branes at the outermost locations $a$ and~$d$ are NS5  branes, while those at~$b$ and~$c$  are D5 branes. We can summarize the corresponding brane configuration in holomorphic supergravity in  the following diagram:

 \begin{equation}
 \begin{tikzcd}[column sep = 0 pt, row sep = 2 pt]
 \text{bulk:}\quad & \C_{z_1} &\times& \C_{w_2}  &\times& \C_{z_2}  &\times& \C_{w_1}  &\times& \C_{z_3}  \\
  &\cong& &\cong& &\cong& &\cong& &\cong& &\\
 & \R^2_{01} &\times& \R^2_{26} &\times& \R^2_{34} &\times& \R^2_{59} &\times& \R^2_{78} \\ \hline
 \text{D3:}\quad & \C_{z_1} &\times& {[a,b]} \times \R_2 &\times& 0 &\times& 0 &\times& 0 \\
  \text{D3:}\quad & \C_{z_1} &\times& {[b,c]} \times \R_2 &\times& 0 &\times& (0\times \alpha) &\times& (\beta\times\gamma) \\
   \text{D3:}\quad & \C_{z_1} &\times& {[c,d]} \times \R_2 &\times& 0 &\times& 0 &\times& 0 \\ \hline
 \text{D5}_b: \quad & \C_{z_1} &\times& b \times \R_2 &\times& 0 &\times& \R_9 &\times& \C_{z_3} \\
 \text{D5}_c: \quad & \C_{z_1} &\times& c \times \R_2 &\times& 0 &\times& \R_9 &\times& \C_{z_3} \\ \hline
 \text{NS5}_a: \quad & \C_{z_1} &\times& a \times \R_2 &\times& \C_{z_2} &\times& \R_5 &\times& 0 \\
  \text{NS5}_d: \quad & \C_{z_1} &\times& d \times \R_2 &\times& \C_{z_2} &\times& \R_5 &\times& 0
 \end{tikzcd}
\end{equation}

It is clear from this table that one holomorphic coordinate (the $z_3  =  \beta +  i \gamma$ coordinate describing the position of the D3 branes that are stretched between D5 branes) survives the $Q_C$ twist and geometrically realizes a holomorphic coordinate on the Higgs branch of the theory.  In the $Q_H$ twist, $z_3$ becomes a topological coordinate, so that the $z_3$ moduli no longer  define nontrivial observables of the $H$-twisted theory.

After the Hanany--Witten transition, we take $d$ to the left of $b$, and correspondingly label it $a'$. Things then look as follows:

 \begin{equation}
 \begin{tikzcd}[column sep = 0 pt, row sep = 2 pt]
 \text{bulk:}\quad & \C_{z_1} &\times& \C_{w_2}  &\times& \C_{z_2}  &\times& \C_{w_1}  &\times& \C_{z_3}  \\
  &\cong& &\cong& &\cong& &\cong& &\cong& &\\
 & \R^2_{01} &\times& \R^2_{26} &\times& \R^2_{34} &\times& \R^2_{59} &\times& \R^2_{78} \\ \hline
 \text{D3:}\quad & \C_{z_1} &\times& {[a,a']} \times \R_2 &\times& (\alpha \times \beta) &\times& (\gamma\times 0) &\times& 0  \\
  \text{D3:}\quad & \C_{z_1} &\times& {[a',b]} \times \R_2 &\times& 0 &\times& 0 &\times& 0 \\
   \text{D3:}\quad & \C_{z_1} &\times& {[a',c]} \times \R_2 &\times& 0 &\times& 0 &\times& 0 \\ \hline
 \text{D5}_b: \quad & \C_{z_1} &\times& b \times \R_2 &\times& 0 &\times& \R_9 &\times& \C_{z_3} \\
 \text{D5}_c: \quad & \C_{z_1} &\times& c \times \R_2 &\times& 0 &\times& \R_9 &\times& \C_{z_3} \\ \hline
 \text{NS5}_a: \quad & \C_{z_1} &\times& a \times \R_2 &\times& \C_{z_2} &\times& \R_5 &\times& 0 \\
  \text{NS5}_{a'}: \quad & \C_{z_1} &\times& a' \times \R_2 &\times& \C_{z_2} &\times& \R_5 &\times& 0
 \end{tikzcd}
\end{equation}

The Coulomb branch of the theory is now geometrically realized via the motion of those D3  branes that extend between NS5 branes; these coordinates (more precisely, the holomorphic coordinate $z_2 = \alpha + i \beta$) survive the $Q_H$ twist and describe the Coulomb branch  of the  theory. In the $Q_C$ twist, $z_2$  becomes a topological direction, so that the positions of the branes are $Q$-exact as moduli of the theory.

\section{Summary of results and future directions}
In this work we have revisited the problem of incorporating boundary conditions of $\N=(0,4)$ type into topological twists of 3d $\N=4$ theories. In previous analyses, \cite{CostelloGaiotto, GaiottoRapcak} the incompatibility of such boundary conditions with the topological twist of the bulk theory was resolved by an appropriate deformation of the boundary conditions, rendering them compatible with bulk supercharges of topological type. One considers a one-parameter family of supercharges that deforms the holomorphic supercharge into a topological one; the deformation of the boundary condition is similarly controlled by the same parameter. Depending on the twist of interest, boundary conditions may or may not admit deformations of this type.

One goal of the present work was to develop a bit more intuition about deformability and non-deformability of such boundary conditions.  We reviewed the holomorphic twists of 3d $\N=4$ multiplets, always working in the perturbative free limit, and described the holomorphic boundary conditions that arise from standard $\N=(0,4)$ and $\N=(2,2)$ boundary conditions explicitly as Lagrangians.
We emphasized that the notion of deformability should imply that the holomorphic boundary conditions are compatible on the nose with the topological twists, when considered as further deformations of the holomorphic theory;  the deformation of the untwisted boundary condition can be thought of as witnessing the $Q_\hol$-exactness of the failure to  be compatible with the deforming supercharge, though we did not  develop this idea in great detail.

We then embedded the construction of the holomorphically twisted theory into type IIB string theory; following~\cite{HananyWitten,Chung:2016pgt,Hanany:2018hlz}, the theory and its boundaries are realized as worldvolume theories of intersecting brane configurations. To holomorphically twist the worldvolume theories, we need to place them in an  appropriate twisted supergravity background in the sense of~\cite{CostelloLi}. We build up the construction of the 3d $\N=4$ theories and their boundary conditions in this setting step by step, and identify the deformations of the supergravity background that induce the further $C$ and~$H$ twists. For D-branes, we find agreement with expectations about deformability at all steps; this leads to a conjectural identification of one deformed boundary condition with an interface between a Lagrangian and a coisotropic brane in the topological $A$-model on~$\C^4$.
We go on to remark on the interpretation of moduli as transverse deformations of branes,  and find it pleasing that the form of the observables in the topologically twisted theory can be seen in terms of the properties of  the corresponding supergravity background---in particular, which  transverse directions become topological and which  remain holomorphic. In this sense, we hope that the  examples worked out  here will serve, not only to build intuition  for the question of  deformability, but also intuition for, and interest in, applications of twisted supergravity in the context of geometric engineering.

There are numerous interesting and important things that we did not do, and which might be worthy of exploration. 
We did not, for example, consider either interactions, nonperturbative questions, or quantization at all; rigorous definitions of twisted 3d $\N=4$ theories at a nonperturbative level and proofs of the Hanany--Witten moves have been given, for example in~\cite{HY}.
We also did not explicitly make connection to the vertex algebras of~\cite{GaiottoRapcak}; this would be nice to do in some concrete examples. It would also be interesting to give a more careful treatment of deformable boundary conditions in the language of derived geometry, and to  derive the new terms  in the action explicitly  or connect to the framework of currents used in~\cite{CostelloGaiotto}.

Although the focus of this paper was on examples in the context of 3d $\N=4$ theories, we emphasize that our interests were more general: holomorphic boundary conditions for topological field theories; producing such boundary conditions via ``deformations'' of supersymmetric boundary conditions of Costello--Gaiotto type; and studying twists of field theories via twisted geometric engineering. 
All of these questions could be studied more broadly. Regarding deformable boundary conditions, the easiest generalization of our setup would be to other odd-dimensional field theories that have sufficient extended supersymmetry to admit a non-minimal twist. Essentially the unique example is five-dimensional $\N=2$ gauge theory. It would be interesting to study boundary conditions that deform in an analogous fashion there. These would likely be four-dimensional holomorphic theories coupled to a holomorphic/topological bulk theory on a geometry of the form $\C \times \R^2 \times \R_+$, or to a fully topological theory. The theory could be embedded in twisted supergravity by considering D4 branes in the $\mathfrak{sl}(4)$-invariant twist of type IIA. We hope to return to such constructions in future work.

Essentially any construction in geometric engineering could be reformulated at a twisted level. Just to give an example, brane configurations realizing 3d $\N=4$ theories with half-BPS defect operators have been analyzed in~\cite{Assel:2015oxa, Assel:2017hck}. An extensive and rigourous field-theoretic analysis of such constructions, in the context of twisting, was given  in~\cite{DGGH}. It would be amusing to set up the brane configurations that engineer these operators directly in twisted supergravity. It would also be interesting to push the interpretation of  the moduli  space of the theory in terms of deformations of the supports of the branes a bit farther. 
Moduli spaces of three-dimensional $\N=4$ theories have been the subject of intense study; see, just for example, \cite{Nakajima:2015txa,Braverman:2016wma,BDG,BDGH,Assel:2017hck,Cremonesi:2014uva,Bachas:2019jaa,HKW}. 
One might hope that twisted supergravity constructions, which allow one to combine twisting directly with the geometric intuition coming from brane setups, will give yet another perspective on this area.

\bibliography{generic}{}
\bibliographystyle{hep}

\end{document}